\documentclass[11pt]{article}

\def\tr{\;{\rm tr}\;}

\def\bra{\langle}   \def\ket{\rangle}
\def\pr{\prime}
\def\implies{\Rightarrow}

\newcommand{\tl}[1]{\tilde{#1}}

\newcommand{\dd}[2]{\frac {\partial #1}{\partial #2}}
\newcommand{\pdr}{\partial}
\newcommand{\grad}{\nabla}

\newcommand{\beq}{\begin{eqnarray}}
\newcommand{\eeq}{\end{eqnarray}}

\newcommand{\half}{\tiny \frac{1}{2}}

\newcommand{\ov}[1]{\frac{1}{#1}}
\newcommand{\fr}[2]{\frac{#1}{#2}}

      \def\gb{\beta}   \def\g{\gamma}       \def\G{\Gamma}
\def\gd{\delta}      \def\D{\Delta}  \def\eps{\epsilon} 
          \def\la{\lambda}      \def\La{\Lambda}
                      \def\vphi{\varphi}
             
     \def\si{\sigma}  \def\Si{\Sigma}     \newcommand{\vsi}{\varsigma}

      \def\Om{\Omega}  

\newcommand{\B}{B} 

\usepackage{hyperref}
\usepackage{graphics}
\usepackage{graphicx}
\usepackage{epsf}
\input{epsf}

\begin{document}

\begin{titlepage}

\title{\normalsize \hfill  
{\sf J.~Phys.~A:~Math.~Theor.~42~(2009)~345403.} \\
\hfill {\tt arXiv:hep-th/0904.4799 } \\ 
\vskip 0mm 
\Large\bf Possible large-$N$ fixed-points and naturalness for $O(N)$ scalar fields}

\author{Govind S. Krishnaswami}
\date{\normalsize Department of Mathematical Sciences \& Centre for Particle Theory, \\ Durham University, Science Site, South Road, Durham, DH1 3LE, UK \vspace{.2in}\\
Chennai Mathematical Institute, \\
Padur PO, Siruseri 603103, India.
\smallskip \\ e-mail: \tt govind.krishnaswami@durham.ac.uk \\ 12 June, 2009}

\maketitle

\begin{quotation} \noindent {\large\bf Abstract } \medskip \\

We try to use scale-invariance and the large-$N$ limit to find a non-trivial $4$d $O(N)$ scalar field model with controlled UV behavior and naturally light scalar excitations. The principle is to fix interactions by requiring the effective action for space-time dependent background fields to be finite and scale-invariant when regulators are removed. We find a line of non-trivial UV fixed-points in the large-$N$ limit, parameterized by a dimensionless coupling. They reduce to classical $\la \phi^4$ theory when $\hbar \to 0$. For $\hbar \ne 0$, neither action nor measure is scale-invariant, but the effective action is. Scale invariance makes it natural to set a mass deformation to zero. The model has phases where $O(N)$ invariance is unbroken or spontaneously broken. Masses of the lightest excitations above the unbroken vacuum are found. We derive a non-linear equation for oscillations about the broken vacuum. The interaction potential is shown to have a locality property at large-$N$. In $3$d, our construction reduces to the line of large-$N$ fixed-points in $|\phi|^6$ theory.

\end{quotation}



\thispagestyle{empty}

\end{titlepage}




\small

\tableofcontents

\clearpage

\normalsize

\section{Introduction}
\label{s-intro}

We investigate the naturalness concept of 't Hooft \cite{tHooft-naturalness} applied to $4$-dimensional $O(N)$ scalar fields. By this dogma, if there is a scalar particle very light compared to the microscopic scale at which the model is superseded, it must be for a good reason e.g., a symmetry. We observe that a non-trivial fixed-point in scalar field theory would be enough to make small masses natural. For, setting masses to zero would restore symmetry under rescaling. We try to realize this scenario by developing an idea of Rajeev \cite{rajeev-non-triv-fp} to find a fixed-point in the large-$N$ limit.

\subsection{Background and motivations}
\label{s-bkgrnd-motivations}

Many discussions of $4$d QFT begin with massive $\la \phi^4$ theory, but it most likely does not have a non-trivial continuum limit \cite{luscher-weisz,kuti-lin-shen,bender-triviality,rigorous-triviality}. So we wish to know if there is a non-trivial $4$d scalar field theory. Another motivation concerns the UV and naturalness problems in the scalar sector of the standard model of particle physics. The importance of  QFTs with controlled UV behavior is well-known: Yang-Mills theories, with a gaussian UV fixed-point provide our best models for strong and weak interactions. In equilibrium statistical mechanics of magnets, the gaussian fixed-point (GFP) controls high energy behavior while the lower energy dynamics is governed by a crossover to the non-trivial Wilson-Fisher fixed-point (WFP)\cite{wilson-kogut}. However, the situation in $4$d massive $\la \phi^4$ theory, the simplest (but unconfirmed) model for W$^{\pm}$,Z masses, is less satisfactory. $\la \phi^4$ is based on the gaussian IR fixed-point, but doesn't flow to any fixed-point in the UV. Perturbatively, interactions become strong (Landau pole) at energies of ${\cal O}(m \exp{[{16\pi^2/3\la}]})$, where $m,\la$ are the parameters of the model in the IR. This is in contrast with asymptotically free theories or theories based on an interacting UV fixed-point which might (in principle) be valid up to a higher energy. Numerical \cite{luscher-weisz,kuti-lin-shen} and analytic\cite{rigorous-triviality,bender-triviality} calculations suggest that without a UV cutoff, $\la \phi^4$ theory is `trivial'\footnote{The renormalized coupling constant vanishes identically and correlations satisfy Wick's formula.}. Unfortunately, the non-trivial WFP in $\phi^4$ theory in $4$-$\eps$ dimensions merges with the GFP in 4d. Pragmatically, lack of a UV fixed-point in $\la \phi^4$ does not prevent us from using it as an effective theory with a cutoff or as a perturbatively defined model like QED, over a range of relatively low energies. Another possibility is that gauge and Yukawa couplings make the standard model better behaved than the scalars in isolation.

Naturalness (appendix  \ref{a-eg-naturalness}) is another problem with $4d$ $\la \phi^4$. In QED, a small electron mass is natural because setting $m_e = 0$ gives QED an additional chiral symmetry, not broken by quantum effects. In massive $\la \phi^4$ theory, classical scale-invariance at $m=0$ is lost quantum mechanically due to a scale anomaly. In the absence of any symmetry, naturalness suggests that the lightest scalar should have a mass $m_H \sim \La$ where $\La$ is the micro-scale at which the model is superseded. But a large $m_{H}$ is problematic. Perturbative unitarity is violated if $m_H$ is too large (at $m_H = \infty$ we return to the theory of massive vector bosons) \cite{pert-unit-bound}. The perturbative unitarity bound from a partial wave analysis of $W$-$Z$ scattering is estimated at $\sim 1$ TeV. Moreover, triviality of the continuum theory implies a `triviality bound' $m_{H} \le {\cal O}(1 ~{\rm TeV})$ \cite{triviality-bound}. There is no non-perturbative cure for these problems, they also arise on the lattice \cite{luscher-weisz,kuti-lin-shen} and in other analytical approaches\cite{bender-triviality,rigorous-triviality}. An advantage of a model with naturally small parameters is that it could in principle be valid over a larger energy range. Indeed, the naturalness breakdown mass scale is estimated at a few hundred GeV for the electroweak standard model\cite{tHooft-naturalness}. 

Another issue is the fine tuning problem: the $1$-loop correction $\la \La^2/ 16 \pi^2$ to $m_H^2$ is quadratic in the momentum cutoff. If a large $\La$ is to be maintained, either the effective $m_H$ is ${\cal O}(\La)$ or $m_H^{\rm bare}$ must be fine-tuned to cancel the radiative correction. We mentioned the difficulties with a large $m_H$. A way out is for $\La$ to be relatively small, but then what replaces $\la \phi^4$ beyond $\La$? One may argue that regulators must be sent to limiting values before making physical conclusions, and that quadratic divergences are absent in some schemes. But without a regulator, $\la \phi^4$ is non-interacting and fails to generate masses. By contrast, non-trivial models based on a UV fixed-point, such as QCD, self-consistently predict low energy behavior irrespective of the physics beyond the standard model and its scale. So it is worth seeking a non-trivial scalar field model based on a UV fixed-point, and a symmetry ensuring naturally light scalars. Moreover, there is a relation between naturalness and fine tuning: radiative corrections are often {\em protected} by the symmetry. Despite criticism of $\la \phi^4$, if a light Higgs is found, we may use the model to predict scattering at relatively low energies. It may turn out to be an effective description of a more intricate framework. Alternatives include supersymmetry (SUSY ensures light scalars \cite{SUSY-higgs}, a challenge is to break it without new naturalness problems), technicolor \cite{technicolor}, little Higgs models \cite{little-higgs}, models based on the Coleman-Weinberg\cite{coleman-weinberg} theme \cite{meissner-nicolai} and others \cite{Bardeen:1995kv,slavnov,halpern-huang}. 

\subsection{Main idea}
\label{s-summary}

We try to avoid the difficulties of $\la \phi^4$ without adding new parameters or degrees of freedom to the standard model (SM), in an approximation where gauge and Yukawa couplings vanish. A possibility is to build a model around a nontrivial UV fixed-point. But existing work (conventional $\eps$-, loop and perturbative expansions, numerics in $m-\la$ plane) does not indicate the presence of one\footnote{Halpern and Huang \cite{halpern-huang} argue there may be potentials for which the GFP is UV. This scenario is quite different from what we propose.}. To find one in $d=4$, it helps to have an expansion parameter. We look for a scale-invariant $O(N)$ model in the $1/N$ expansion (for a review of large-$N$ vector models see \cite{zinn-moshe-large-N-review}). An interesting case is $N = 4$ (the scalar sector of SM is $O(4)$ symmetric, broken to custodial $O(3)$ symmetry by the scalar vev). There is a precedent for this. $3$d quantum $\la |\phi|^6$ theory is scale-invariant at $N=\infty$ for any $\la$, though whether there is any non-trivial fixed-point for finite $N$ is unclear \cite{Townsend-phi-sixth,phi-sixth-3d}. In $d=4$, an idea to use the $N \to \infty$ limit to construct a scale-invariant model was given by Rajeev \cite{rajeev-non-triv-fp}. We give up thinking of a QFT as defined by a pre-specified classical action $S$. $S$ is often a useful concept since it approximates the quantum effective action $\Gamma$ as $\hbar \to 0$, where fluctuations from the path integral measure are suppressed. By contrast, in the large-$N$ limit, both `action' and quantum fluctuations from the `measure' are comparable. We pick a non-scale-invariant action to cancel the `scale anomaly' from quantum fluctuations. Strictly, both action and measure are infinite prior to regularization and neither is scale-invariant if regulated. Combined, they produce a finite $1$-parameter($\la$) family of scale-invariant $\Gamma$'s when regulators are removed. $\Gamma$ is physical and defines the theory. $\la$ is the dimensionless coupling of a $\phi^4$-type term, which is marginally irrelevant near the GFP, but whose $\beta$-function vanishes in the large-$N$ limit when considered around the non-trivial fixed-point. Thus the Landau pole of usual $\phi^4$ theory is avoided. An advantage of scale-invariance over SUSY is that it is easy to break by adding the most relevant deformation, a mass term, without introducing new naturalness problems. Setting $m=0$ (for any $\la$) is natural, we gain scale-invariance by doing so. This line of scale-invariant theories are UV with respect to the mass term, and thus ensure controlled UV behavior. For naturally light scalars via scaling symmetry, it suffices to have one fixed-point. It could be that upon including $1/N$ corrections, scale-invariance can be maintained only for one\footnote{There may be no non-trivial fixed-point when $1/N$ corrections are incorporated, then our scenario would fail.} $\la = \la_0$. This would be acceptable, since $m=0, \la = \la_0$ would be natural due to scale-invariance at that point. Our model has only two free parameters at large-$N$, ensuring predictive power. Here we address issues relevant to the UV and naturalness problems of scalars at $N=\infty$, postponing analysis of corrections and renormalizability at finite $N$ and coupling to fermions/gauge fields to future work.



\section{Lagrangian and change of field variables}
\label{s-lagrangian-change-of-var}

Consider a $4$d $N+1$ component Euclidean real scalar field $\phi_{0 \leq i \leq N}$, with a globally $O(N+1)$ invariant action. The factors in the partition function are chosen to facilitate $N \to \infty$
    \beq
    Z = \int [D\phi] \exp{[ -(2\hbar)^{-1} \int d^4x ~ \{|\grad \phi_i|^2 + N
        V({|\phi|^2 / N})  \} ~]}.
    \label{e-original-partition-function}
    \eeq
We introduce the Hubbard-Stratonovich field $\si$ via a Laplace transform with respect to $\eta = {\phi_i \phi_i / N}$. This leaves the action quadratic in $\phi_i$ so that we can integrate them out. 
    \beq
        Z &=& \int [D\phi] \int_0^\infty [D\eta] e^{-\ov{2\hbar} \int d^4x
        \{ |\grad \phi|^2 + NV(\eta) \} } \prod_x \delta(\eta(x) -
        \phi^2(x)/N), \cr 
    \prod_x \delta(N\eta - \phi^2) &=& \prod_x (4 \pi i \hbar)^{-1}~ \int_{\cal C} [D\si]~ e^{\int d^4x ~\si(N\eta - \phi^2)/2\hbar}.
    \eeq
$\cal C$ is any contour from $-i\infty$ to $i\infty$ since the integrand is entire. Up to normalization,
    \beq
    Z = \int [D\phi] \int_0^\infty [D\eta]
        \int_{\cal C} [D\si] e^{-\ov{2\hbar} \int d^4x
        \{ |\grad \phi|^2 + \si \phi^2 + N V(\eta) - N \si \eta \}
        }.
    \eeq
$\si(x)$ is Laplace conjugate to the $O(N+1)$ singlet $\eta$, and can be regarded as a Lagrange multiplier enforcing $\eta = \phi^2/N$. Let $b = \phi_0/\sqrt{N}$ and here on, $[D\phi]$ does not include $\phi_0$:
 	\beq
    Z = \int [Db] \int [D\phi] \int_0^\infty [D\eta] \int_{\cal C} [D\si]
        e^{-\ov{2\hbar} \int d^4x
        \bigg[ \sum_{i=1}^N \{(\grad \phi_i)^2 + \si \phi_i^2 \}
        + N (\grad b)^2 + N \si b^2 + N V(\eta) - N \si \eta
        \bigg]}.
    \eeq
Reverse the $\eta$ and $\si$ integrals and observe that the $\eta$ integral is a Laplace transform at each $x$,
    \beq
    && \int_0^\infty [D\eta] e^{-(N / 2\hbar) \int d^4x [ V(\eta) - \si \eta]}
        = e^{-({N / 2\hbar}) \int d^4x W(\si)}, {\rm ~~~ so ~ that}
    \label{e-V-to-W-laplace transform} \cr
    Z &=& \int [Db] \int [D\phi] \int_{\cal C} [D\si]
    e^{-(1/2\hbar) \int d^4x \bigg[ \sum_{i=1}^N ((\grad \phi_i)^2 + \si
    \phi_i^2) + N (\grad b)^2 + N \si b^2 + N W(\si)  \bigg]}.
    \eeq
Reversal of $\si$ and $\eta$ integrals works if $W$ is non-singular on $\cal C$. Now reverse the $\phi$ and $\si$ integrals. The gaussian $\phi$ integral converges if eigenvalues of $-\grad^2 + \si$ have positive real part:
    \beq
    \int [D \phi] e^{-(1/2\hbar) \sum_{i=1}^N \int d^4x \phi_i (-\grad^2+ \si)
    \phi_i}  = \bigg[ \det{\bigg( \fr{-\grad^2 + \si}{2\pi \hbar}\bigg)}
    \bigg]^{-N/2}.
    \eeq
This is ensured if $\Re \si > 0$ (the answer has an analytic continuation to $\si \in {\bf C \setminus R}^-$). The $\log$ of this determinant needs a scale $M$ for its definition. Thus, up to a multiplicative constant
    \beq
    Z &=& \int_{-\infty}^\infty [Db] \int_{\cal C} [D\si] e^{-N S(b,\si)} \cr
    {\rm where ~~~~} S(b,\si) &=& (2\hbar)^{-1} [\hbar
    \tr \log \{(-\grad^2 + \si) / M^2 \} + \int d^4x~ \{ (\grad b)^2
        + \si b^2 + W(\si) \} ].
    \label{e-action}
    \eeq
A reason to use $\si$ instead of $\phi_i$ is that as $N \to \infty$ holding $\hbar \ne 0$ fixed, $\si$ has small fluctuations, while $\phi_i$ have large fluctuations. 
$\si$ is a dynamical field with self-interactions specified by $W(\si)$. $\si$ is {\em not} the massive $\si$ particle of the symmetry broken $O(N)$ linear sigma model. $\si(x)$ is valued on a contour $\cal C$ from $-i\infty$ to $i\infty$ lying to the right of singularities of $W(\si)$. The contour of integration for $b$ is $\bf R$. Note that $[\si] = {\rm mass}^2$ while $[b] = {\rm mass}$. We kept $b$ since a vev for $b$ signals breaking of $O(N+1)$ invariance; $\sigma$ could acquire a vev without breaking $O(N+1)$. 

\section{Scale-invariance of the effective action at $N=\infty$}
\label{s-scale-inv-of-effac-inf-N}

Our model is built by requiring scale-invariance of the effective action for {\em arbitrary} backgrounds at each order in $1/N$. The interaction $W(\si)$ appearing in the action (\ref{e-action}) is expanded in $1/N$
    \beq
        W(\si) = W_0(\si) + W_1(\si)/N + W_2(\si)/N^2 + \cdots.
    \eeq
$W(\si)$ is {\em not} assumed analytic at $\si = 0$. The action is also expanded in powers of $1/N$
    \beq
        S(b,\si) &=& \hbar^{-1} [S_0 + S_1/N + S_2/N^2 + \cdots] \cr
    {\rm where~~~}
        S_0 &=& (1/2) \bigg[\hbar \tr \log[{(-\grad^2 + \si) / M^2}] +
            \int d^4x \bigg\{(\grad b)^2 + \si b^2 + W_0(\si) \bigg\}
            \bigg], \cr
        S_n &=& (1/2) \int d^4x W_n(\si), ~~~ {\rm for}~~ n =1,2,3,\ldots.
    \eeq
`Counter-terms' $W_{1,2\ldots}(\si,\hbar)$ are chosen to cancel divergences and scale anomalies from fluctuations in $b$ and $\si$ while $W_0$ is chosen to cancel those from fluctuations in $\phi_{1 \ldots N}$. The possible choice(s) of $W_{0,1,2\ldots}$ define the scale-invariant fixed-point(s) just as $\half |\pdr \phi|^2$ defines the trivial fixed-point. $W_n(\sigma)$ are unrestricted, but for predictive power, they can depend on at most a few free parameters. $N$ and $\hbar$ appear differently in $S(b,\si)$. As $N \to \infty$, $b,\si$ have small fluctuations and are governed by the action $S_0(b,\si)$. As $\hbar \to 0$, $\phi_i$ have small fluctuations, they are governed by the action $\int d^4x [|\grad \phi|^2 + N V(\phi^2/N)]$. These two `classical' limits capture different features of the quantum theory for given $W(\si)$. $\hbar \tr\log[(-\grad^2 + \si)/M^2]$ is a quantum correction to the action as $\hbar \to 0$, but part of the `classical' action as $N \to \infty$! 

A theory is scale-invariant if its quantum effective action $\Gamma$ (Legendre transform of generator of connected correlations \cite{jackiw-eff-action,nair-QFT}) is scale-invariant. $\Gamma$ is defined implicitly by
    \beq
    e^{- N \Gamma(\B,\Si)} = \int [D\beta] \int_{\cal C}
        [D\vsi] \exp\bigg[-N \bigg\{S(\B+\beta,\Si+\vsi) 
        - \beta \fr{\delta \Gamma}{\delta \B}
         - \vsi \fr{\delta \Gamma}{\delta \Si} \bigg\} \bigg].
    \label{e-implicit-def-of-eff-action}
    \eeq
$\B(x)$ and $\Si(x)$ are background fields while $\beta$ and $\vsi$(`varsigma') are fluctuating fields, $b = \B + \beta, ~~ \si = \Si + \vsi$. Holding $\hbar$ fixed, $\Gamma$ is expanded as $\Gamma_0 + {\Gamma_1 / N} + {\Gamma_2 / N^2} + \cdots$. From (\ref{e-implicit-def-of-eff-action}),
    \beq
     \Gamma_0 = S_0 = \half [\hbar \tr \log(-\grad^2 + \Si) / M^2
            + \int d^4x \{(\grad \B)^2 + \Si \B^2 + W_0(\Si) \} ].
    \label{e-def-of-eff-ac-at-infinite-N}
    \eeq
${\rm Tr} \log[(-\grad^2 + \Si(x))/M^2]$ is divergent and must be regulated. $W_0(\Si,M,{\rm regulator})$ is chosen so that when the regulator is removed, $\Gamma_0$ is finite and scale-invariant. $W_1$ is found by the same principle applied to $\Gamma_1$ and so on. $\hbar =0$ is a limiting case where there are no quantum fluctuations of $\phi_{1 \ldots N}$ to contribute any divergences or scale violations and the finite $W_0 = -\Si^2/\la$ is the general choice for which $\Gamma_0^{\hbar =0} = \half \int d^4x \{ (\grad B)^2 + \Si B^2 - \Si^2/\la \}$ is scale-invariant for any $\la$, corresponding to a quartic original potential $V(\eta) = \la\eta^2/4$.

\subsection{Effective action for constant background $\Si(x)=\Si_o$ at $N = \infty$}
\label{s-cl-eff-ac-const-bkgrnd}

${\rm Tr} \log[{(-\grad^2 + \Si(x)) / M^2}]$ (\ref{e-def-of-eff-ac-at-infinite-N}) is easily found for a constant $\Si$ so consider this case first. 

\subsubsection{Momentum cutoff regularization}
\label{s-mom-cut-off-regularization}

In momentum cutoff regularization ($\int d^4x = \Om$, $\int d\Om_4 = 2\pi^2$ is the `area' of $S^3$)
    \beq
    \hbar \tr \log{-\grad^2 + \Si_o \over M^2} 
    &=& [{2\pi^2 \hbar \Om / (2\pi)^4}]~ \int_0^\La dp~ p^3 
    	~ \log{[(p^2 + \Si_o)/ M^2]} \cr
    &=& {\hbar \Om \over 64 \pi^2} \bigg[2 \La^4 \log{\La^2 + \Si_o \over M^2} 
    	- \La^4 + 2 \La^2 \Si_o 
    	- 2 \Si_o^2 \log{\La^2 + \Si_o \over \Si_o} \bigg] \cr
    &=& {\hbar \Om \over 64 \pi^2} \bigg[ 2 \La^4 \log{\La^2 \over M^2} -
        \La^4 + 4 \La^2 \Si_o - 2 \Si_o^2 \log{\La^2 \over M^2} ~~({\rm divergent}) 	\cr
    && + 2 \Si_o^2 \log(\Si_o/M^2) ~~({\rm scale~violating~finite}) \cr
    && - \Si_o^2 ~~({\rm scale~invariant~ finite})
    + {\rm ~terms~vanishing~ as~} \La \to \infty \bigg].
    \eeq
We must pick $W_0(\Si)$ such that (\ref{e-def-of-eff-ac-at-infinite-N}) is finite and scale-invariant when $\La \to \infty$. In sec. \ref{s-cl-eff-ac-arbit-bkgrnd} we do this for general $\Si,\B$. Here we get an idea of the answer by requiring that $\Gamma_0(\B,\Si)$ be scale-invariant for constant $\Si = \Si_o$. Hence, pick the `minimal subtraction' choice
    \beq
    W_0^\La(\Si_o)  = {- \hbar \over 64 \pi^2} \bigg[2 \La^4 \log{\La^2 \over M^2}
        - \La^4 + 4 \La^2 \Si_o - 2 \Si_o^2 \log{\La^2 \over M^2} \bigg]
        - {\hbar \over 32 \pi^2} \Si_o^2 \log{\Si_o \over M^2}.
    \eeq 
Of course, we could add a finite term $-\Si_o^2/\la$ to $W_0(\Si_o)$ and preserve scale-invariance of $\G_0$. So at $N = \infty$, there is a $1$-parameter($\la$) family of Renormalization Group (RG) fixed-points. Adding $m^2 \Si_o$ is a relevant deformation while $c_n \Si_o^n$ for $n > 2$ are irrelevant, as $c_n$ have negative mass dimensions. So consider the mass deformed theory, where $W_0(\Si_o)$ is a $2$-parameter $(m,\la)$ family
    \beq
    W_0^\La(\Si_o)  = {- \hbar \over 64 \pi^2} 
    	\bigg[2 \La^4 \log{\La^2 \over M^2}
        - \La^4 + 4 \La^2 \Si_o - 2 \Si_o^2 \log{\La^2 \over M^2} \bigg]
        - {\hbar \Si_o^2 \over 32 \pi^2}  \log{ \Si_o \over M^2} 
        - \Si_o^2/\la + m^2 \Si_o.
    \eeq
$W_0$ has a cut along ${\bf R}^-$, so the contour $\cal C$ in (\ref{e-implicit-def-of-eff-action}) must miss ${\bf R}^-$. The corresponding $\Gamma_0$ (\ref{e-def-of-eff-ac-at-infinite-N}) is independent of $M$:
    \beq
    \G_0(B_o,\Si_o) = (\Om / 2)~ [ m^2 \Si_o - (\la^{-1} + {\hbar / 64 \pi^2}) \Si_o^2 + \Si_o B^2 ].
    \eeq
These fixed-points are UV with respect to $m^2 \Sigma_o$. Recall that the massless free field $|\grad \phi|^2$ is UV with respect to $m^2 \phi^2$, but IR with respect to $\la \phi^4$. In our model, the analogue of the quartic coupling, $- \Si_o^2/\la$, is exactly marginal. So for any $\la, \hbar$, we can set $m=0$ and gain scale-invariance: $m$ can be naturally small. Though some formulae are familiar from the Coleman-Weinberg calculation \cite{coleman-weinberg}, the physical principles and interpretation are quite different. While they tried to generate masses through quantum corrections to classical massless $\phi^4$ theory, our aim is to find a {\it different} theory that is quantum mechanically scale-invariant.

\subsubsection{Analytic (Zeta function) regularization}
\label{s-zeta-fn-regularization}

We recalculate $\Gamma_0$ by $\zeta$-regularizing $\tr \log[(-\grad^2 + \Si_o)/M^2]$, which directly prescribes a scale-violating finite part for $\tr \log[(-\grad^2 + \Si_o)/M^2]$. We pick $W_0(\Si_o)$ to cancel this, so that $\Gamma_0(B,\Si_o)$ is scale-invariant and finite. In $\zeta$-regularization, we do not prescribe how $W_0(\Si_o)$ depends on a regulator. Such a short-cut is not possible in other schemes. We will often use $\zeta$-regularization, but comparison of different schemes allows us to identify scheme dependence. The $1$-parameter family of fixed-points exist independent of scheme and the effective potential is also independent up to a finite shift in $1/\la$. Let
    \beq
    \zeta(s) = \tr \bigg[{(-\grad^2 + \Si_o) \over M^2} \bigg]^{-s}
    = \Om \int \fr{d^4p}{(2\pi)^4}~ {M^{2s} \over [p^2 + \Si_o]^s}.
    \eeq
$\zeta(s)$ is clearly analytic for $\Re{s} > 2$, and in fact is meromorphic with simple poles at $s=1,2$ 
    \beq
        \fr{\zeta(s)}{\Om} = {M^{2s} \over (2\pi)^4} \int d\Om_4 
        \int_0^\infty \fr{p^3 dp}{(p^2 + \Si_o)^s}
        = {M^{2s} \over 16 \pi^2} \fr{\Si_o^{2-s}}{(s-1)(s-2)}.
    \label{e-zeta-of-s-const-bkgrnd}
    \eeq
In particular, $\zeta(s)$ is regular at $s=0$ and may be used to define 
    \beq
    \tr \log[(-\grad^2 + \Si_o)/M^2] ~\equiv~ -\zeta^\pr(0) ~=~
    (\Si_o^2 \Om / 32 \pi^2) ~\log[{e^{-3/2} \Si_o/M^2}].
    \label{e-tr-log-const-bkgrnd}
    \eeq
So the effective action at $N=\infty$ for constant backgrounds is
    \beq
        \Gamma_0(B_o,\Si_o) = (\Om / 2) \bigg[ (\hbar \Si_o^2 / 32 \pi^2)
        \log[e^{-3/2} {\Si_o / M^2}] + \Si_o B_o^2 + W_0(\Si_o) \bigg].
    \label{e-eff-ac-at-inf-N-const-bkgrnd}
    \eeq
The choice of $W_0(\si)$ that ensures $\Gamma_0(\B,\Si_o)$ is scale-free (for $m=0$) is
    \beq
    W_0(\si_o) = m^2 \si_o - \si_o^2/\la - (\hbar \si_o^2 / 32 \pi^2)
        \log[e^{-3/2} {\si_o / M^2}].
    \label{e-W_0-for-constant-bkgrnd}
    \eeq
We added a relevant mass perturbation away from the line of fixed-points parameterized by $\la$. The terms in $W_0$ are of different orders in $\hbar$ but all of order $N^0$. For this choice of $W_0$, we get
    \beq
    \Gamma_0(\B_o,\Si_o) = (\Om / 2) ~[m^2 \Si_o
    - \Si_o^2/\la + \Si_o \B_o^2 ].
    \label{e-gamma0-zeta-fn-const-bkgrnd}
    \eeq
$M$ cancels out from the effective potential, which is scale-free for $m=0$. Though $W_0(\si_o)$ has a cut along ${\bf R}^-$, $\Gamma_0$ and $S_0$ are entire for constant backgrounds. Comparing with sec. \ref{s-mom-cut-off-regularization} we see that independent of regularization scheme, there is a $1$-parameter family of fixed-points labelled by $\la$. But $\lambda$ itself is scheme dependent,
    \beq
    \la_{\zeta}^{-1} = \la_{\rm cutoff}^{-1} + {\hbar / 64 \pi^2}  
    	~~~ {\rm as}~~ N \to \infty.
	\label{e-scheme-dependent-lambda}
    \eeq

\subsection{$N = \infty$ effective action expanded around a constant background}
\label{s-cl-eff-ac-arbit-bkgrnd}

Here, we expand $\Gamma_0(\B,\Si)$ (\ref{e-def-of-eff-ac-at-infinite-N}) in powers and derivatives of $\vsi / \Si_o$ where $\Si_o \ne 0$ is a constant background and $\vsi(x) = \Si(x) - \Si_o$. From appendix \ref{a-exp-for-tr-log} (\ref{e-tr-log-ignoring-deriv-of-cubic-and-higher}), in $\zeta$-regularization,
    \beq
    \tr\log{\Si -\grad^2 \over M^2} = {\Si_o^2 \Om \over 32\pi^2}
        \log{\Si_o e^{-3/2} \over M^2} + \int {d^4x \over 16\pi^2} \bigg[
        \Si_o \vsi \log{\Si_o \over M^2 e} 
        + {\vsi^2 \over 2} \log{\Si_o \over M^2}
            - \vsi \{ \Pi(\Delta) + \Pi({\vsi \over \Si_o}) \} \vsi  \bigg]
    \label{e-tr-log-arbit-bkgrnd}
    \eeq
up to cubic/higher order terms in $\vsi$ also involving gradients i.e. ${\cal O}(\vsi^3,\grad^2)$. We focus on these terms as they are the ones needed to study small oscillations around extrema of the effective action. This reduces to (\ref{e-tr-log-const-bkgrnd}) for $\Si= \Si_o$. The remaining terms follow by the method of appendix \ref{a-exp-for-tr-log}, but are independent of the scale $M$. Here $\Si_o$ is arbitrary and need not be the average value of $\Si$, which may be $0$. In sec. \ref{s-cancel-scale-anomaly} we show that the scale dependent part of $\tr \log[(-\grad^2 + \Si(x))/M^2]$ is restricted to the first $3$ terms on the rhs of (\ref{e-tr-log-arbit-bkgrnd}). Note that $\Delta = -\grad^2
/\Si_o$ and
    \beq
        \Pi(\Delta) = \sum_{n=1}^\infty {(-\Delta)^{n}
            \over n(n+1)(n+2)} = {\Delta(3 \Delta +2) -2(\Delta+1)^2
            \log{(1+\Delta)} \over 4 \Delta^2}
        = -{\Delta \over 6} + {\Delta^2 \over 24} - {\Delta^3
            \over 60} + \cdots
    \eeq
Thus, the effective action in $\zeta$-regularization at $N=\infty$ is
    \beq
        2~ \Gamma_0(\B,\Si) &=& \int d^4x \bigg[(\grad \B)^2 + \si \B^2 + W_0(\Si) 		+ \fr{\hbar}{16\pi^2}
        \bigg\{\half \Si_o^2 \log{\Si_o e^{-3/2} \over M^2} + \Si_o
         \vsi \log{\Si_o \over eM^2} \cr && 
         + (\vsi^2 /2) \log(\Si_o / M^2) - \vsi \{ \Pi(\Delta) 
         + \Pi(\vsi/\Si_o) \} \vsi + {\cal O}(\vsi^3,\grad)
         \bigg\} \bigg].
    \label{e-eff-ac-large-N-arbit-bkgrnd}
    \eeq

\subsection{Fixing the interaction at $N=\infty$ by requiring scale-invariance}
\label{s-fix-W_0-arbit-bkgrnd}

The $\zeta$-regularized choice of $W_0$ that makes $\Gamma_0$ (\ref{e-eff-ac-large-N-arbit-bkgrnd}) scale-free for $m=0$ is ($\vsi = \si - \si_o$)
    \beq
        W_0(\si) = m^2 \si - {\si^2 \over \la} - \fr{\hbar}{16\pi^2}
        \bigg\{ {\si_o^2 \over 2} \log{\si_o e^{-{3 / 2}} \over M^2} 
        + \si_o \vsi \log{\si_o \over M^2 e} 
        + {\vsi^2 \over 2} \log{\si_o \over M^2} \bigg\}
    \label{e-W_0}
    \eeq
For $m \ne 0$ we have a mass deformation. Notice that the finite part of $W_0$ in $\zeta$-regularization is a {\em local} function of $\sigma(x)$. We will show in sec.\ref{s-locality-of-W} that the divergent part of $W_0(\sigma)$, which is suppressed in $\zeta$-regularization, is also local. $M$ appears in $W_0$, but cancels out in the $N=\infty$ effective action which is now finite (for $\Si_o \ne 0$)
    \beq
        2 ~ \Gamma_0 = \int {d^4x} [(\grad \B)^2 + \Si \B^2 + m^2 \Si 
        - \Si^2/\la - (\hbar / 16\pi^2)
            \{\vsi \{ \Pi(\Delta) + \Pi(\vsi/\Si_o) \} \vsi 
            + {\cal O}(\vsi^3,\grad^2) \} ].
    \label{e-eff-ac-large-N-finite-and-scale-invariant}
    \eeq
$\Si_o$ is arbitrary, it is {\em not} a free parameter. It appears merely because we study the theory around a constant background. $\Gamma_0(B,\Si)$ has interactions in the absence of regulators, indeed it has an infinite number of proper vertices. This indicates our theory is not trivial at $N = \infty$. As before, the constant $\Si_o \ne 0$ is arbitrary. The two terms $\propto \hbar$ are precisely the ones needed to study long wavelength small oscillations around the $O(N+1)$ symmetric and broken extrema of $\Gamma_0$ in sec. \ref{s-small-osc}. The two derivative term in $\vsi \Pi(\D) \vsi$ contributes around the symmetric vacuum. Though $\vsi^2 \Pi(\vsi/\Si_o) = -\vsi^3/6\Si_o + \vsi^4/24 \Si_o^2 -\ldots$ is at least cubic in $\vsi$, it contains no derivatives and so is important for long wavelength oscillations, especially when $\Si_o$ is small as in the broken phase. The omitted ${\cal O}(\vsi^3,\grad^2)$ and higher order terms in $\vsi/\Si_o$ are all finite, scale-free and calculable by the method of appendix \ref{a-exp-for-tr-log}, but they do not contribute to long wavelength small oscillations. 

\subsection{Cancelation of scale anomaly}
\label{s-cancel-scale-anomaly}

Under dilations, quantities are canonically rescaled 
    \beq
        D_a x^\mu = a^{-1} x^\mu, ~~ D_a b = a b, ~~ D_a \si = a^2 \si,
        ~~ D_a \la = \la.
    \label{e-scale-transformation}
    \eeq
If we also rescaled the physical scales $m$ and $M$, dilation invariance would be vacuous. The generator of infinitesimal dilations is defined as $\delta_D = \lim_{\eps \to 0} \{D_{1+\eps}  - {\bf 1} \}/\eps$, so that
	\beq
        \delta_D x^\mu &=& - x^\mu, ~~ \delta_D b = b, ~~ \delta_D \si = 2 \si,
        ~~ \delta_D \la = 0 \cr
       {\rm and~~} \delta_D &=& -x^\mu \dd{}{x^\mu} + b(x) \dd{}{b(x)} 
       + 2 \si(x) \dd{}{\si(x)}.
	\label{e-dilatation-generator}
    \eeq
We show that $\Gamma_0(B(x),\Si(x))$ (\ref{e-def-of-eff-ac-at-infinite-N}) is dilation invariant if $W_0(\Si)$ is as in (\ref{e-W_0}) with $m=0$. First, $\int d^4x \{ (\grad B)^2 + \Si B^2 \}$ is unchanged under $D_a$, so $W_0(\Si)$ and $\tr \log[(-\grad^2 + \Si)/M^2]$ are the only terms in (\ref{e-def-of-eff-ac-at-infinite-N}) with non-trivial (in fact inhomogeneous) dilations.
    \beq
    D_a \int d^4x W_0(\Si) &=& \int d^4x W_0(\Si) - \fr{\hbar}{16\pi^2} \int d^4x
        \bigg[ \half \Si_o^2 \log{a^2}
        + \Si_o \vsi \log{a^2}  + {\vsi^2 \over 2} \log{a^2}
        \bigg] \cr
    &=& \int d^4x W_0(\Si) - \fr{\hbar \Si_o^2 \Om \log a}{8\pi^2}
        \bigg[\half + \fr{\bra \vsi \ket}{\Si_o} 
        + \fr{\bra \vsi^2 \ket}{2 \Si_o^2} \bigg]
    \cr
    \Rightarrow ~~ \delta_D \int d^4x W_0(\Si) &=& - \fr{\hbar \Om \Si_o^2}{8
        \pi^2} \bigg[\half + \fr{\bra \vsi \ket}{\Si_o} + \fr{\bra \vsi^2 \ket}{2 \Si_o^2} \bigg]
        \label{e-scale-tr-of-W0}
    \eeq
where $\vsi = \Si - \Si_o$ and $\Si_o$ is a constant background. If $\zeta(s) = \tr [-\grad^2 + \Si]^{-s}$, then $\tr \log[(-\grad^2 + \Si)/M^2] = -\zeta^\pr(0) - \zeta(0) \log{M^2}$, where $\zeta(0)$ is scale-invariant (appendix \ref{a-scale-anomaly}). Now $D_a \zeta(s) = a^{-2s} \zeta(s)$ $\implies$ $D_a \zeta^{\prime}(s) = - 2 \zeta(s) a^{-2s}\log a + a^{-2s} \zeta^{\prime}(s)$, whence $D_a \zeta^{\prime}(0) = \zeta^{\prime}(0) - 2 \zeta(0) \log a$. So, $\delta_D \zeta^\prime(0) = -2 \zeta(0)$. From appendix (\ref{a-scale-anomaly})
    \beq
        \delta_D \hbar \tr \log[{-\grad^2 + \Si \over M^2}] = - \hbar \delta_D
        \zeta^\prime(0) = 2 \hbar \zeta(0) 
        = {\hbar \Om \Si_o^2 \over 8 \pi^2} \bigg[
            \half + {\bra \vsi \ket \over \Si_o} 
            + {\bra \vsi^2 \ket \over 2 \Si_o^2}
            \bigg].
    \label{e-scale-tr-of-tr-log}
    \eeq
We see that the scale anomaly of $\tr\log[(-\grad^2 + \Si) / M^2]$ (\ref{e-scale-tr-of-tr-log}) exactly cancels that of $\int d^4x W_0(\Si)$ (\ref{e-scale-tr-of-W0}). So for $m=0$, $\Gamma_0$ (\ref{e-eff-ac-large-N-finite-and-scale-invariant})
is dilation invariant: $\delta_D \Gamma_0(B,\Si) = 0$.
However, with a mass term,
    \beq
    \gd_D \G_0(B,\Si) = -2 \int d^4x~ m^2~ \Si(x) = -m \dd{}{m} \int d^4x~
    m^2~  \Si(x).
    \eeq
Define  $\beta_0^m = m,  \beta_0^\la =0, \gamma_0^b = 1, \gamma_0^\sigma = 2$ and the large-$N$ renormalization group vector field 
	\beq
	\gd_0 = -x^\mu \pdr_\mu + \gamma_0^b b(x) \dd{}{b(x)} 
		+ \gamma_0^\sigma \sigma(x) \dd{}{\sigma(x)} 
		+ 	\beta_0^m \dd{}{m} + \beta_0^\la \dd{}{\la} 
	\label{e-RGE-vfld-large-N}
	\eeq    
Then $\Gamma_0$ satisfies the RG equation $\gd_0 \G_0 =0$. Both $\Gamma$ and $\gd$ may receive $1/N$ corrections while satisfying the RGE $\gd \G = 0$. But for a fixed-point, we need $\gb^\la =0$ for at least one $\la$ at each order in $1/N$.

\section{Locality of interaction potential $W(\sigma(x))$}
\label{s-locality-of-W}





We observed in $\zeta$-regularization (sec. \ref{s-fix-W_0-arbit-bkgrnd}) that the scale-violating part of $W_0(\sigma)$ is local in $\si(x)$. What about the divergent counter terms? As $N \to \infty$, quantum fluctuations contribute $\hbar \tr \log [(-\grad^2 + \Si(x))/M^2]$ to the effective action, and $W_0(\si)$ is chosen to cancel its divergent and scale-violating parts. Here we isolate these parts and show they are local in $\sigma(x)$. Begin with $y^{-1} = \int_0^\infty dt ~e^{-ty}$ where $t$ is an auxiliary `time' variable. Integrate $y$ from $x_0$ to $x$ and assume that the order of integrals can be reversed on the RHS. This gives
    \beq
    \log{x/x_0} = - \int_0^\infty (e^{-tx} - e^{-tx_0}) ~t^{-1}~ dt.
    \eeq
Replacing $x$ by a positive operator $A$ and $x_0$ by a scalar operator $M^2 >0$ and taking a trace (assuming $\tr$ commutes with the integral over $t$, whose dimensions are (length)$^2$)
    \beq
    \tr \log {A/M^2} = - \int_0^\infty  \tr ({e^{-t A} - e^{-t M^2}}) ~t^{-1}~ dt .
    \eeq
$M$ is a parameter with dimensions of mass. Now take $A = -\grad^2 + \sigma(x)$ and use the result $\tr e^{-tA} = \int d^4x ~ h_t(x,x)$ from appendix \ref{a-exp-for-tr-log}. The heat kernel has the expansion ($\si(x) = \si_o + \varsigma(x)$)
    \beq
    h_t(x,x) = {e^{-\si_o t} \over (4 \pi t)^2} \sum_{n=0}^\infty a_n(x,x) t^n
    ~~~~ \implies ~~~~
    \tr e^{-tA} = \Om {e^{-\si_o t} \over (4 \pi t)^2} \sum_{n=0}^\infty \bra a_n(x,x) \ket t^n.
    \eeq
$a_n$ are finite and depend on $\varsigma$ and at most $2n-2$ of its derivatives. $\bra f \ket = \ov{\Om} \int d^4 x f$. The sum on $n$ is often asymptotic; we hope this does not affect our conclusions. Under these hypotheses,
    \beq
    \tr \log {A / M^2} = - \Om \int_0^\infty  \bigg[e^{-\si_o t} (4 \pi t)^{-2} (\bra a_0 \ket + \bra a_1 \ket t + \bra a_2 \ket t^2 + \cdots ) - e^{-t M^2} \bigg] t^{-1} dt
    \eeq
Integrating term by term, we write
    \beq
    \tr \log[(-\grad^2 + \sigma(x)) / M^2] &=& 
    	\Om (T_0 + T_1 + T_2 + T_3 + \ldots) {\rm ~~~~where~~} \cr
    T_0 = - {1 \over 16 \pi^2} \int_0^\infty dt {e^{- \si_o t} \over t^3}, &&
    T_2 = - \int_0^\infty {dt \over t} 
    	\bigg[{e^{- \si_o t} \bra \vsi^2 \ket \over 32 \pi^2} 
		- e^{-M^2 t}  \bigg], \cr
    T_1 = {\bra \vsi \ket \over 16 \pi^2 } 
    	\int_0^\infty dt {e^{-\si_o t} \over t^2}, &&
    T_{n \geq 3} = - {\bra a_n \ket \over 16 \pi^2} 
    	\int_0^\infty dt ~ t^{n-3} e^{- \si_o t} 
		= - {(n-3)! \over 16 \pi^2} {\bra a_n \ket \over \si_o^{n-2}}.
    \eeq
$T_{0,1,2}$ are UV divergent (i.e. as $t \to 0$) while $T_{n \geq 3}$ are finite and scale$(M)$ free. Thus all the divergences and scale-violations are in $T_{0,1,2}$, which we evaluate with a UV cutoff at $t=\ov{\La^2}$
    \beq
    T_0 = - {\La^4 \over 16 \pi^2} E_3(\si_o/\La^2), ~~~
    T_1 = {\bra \vsi \ket \La^2 \over 16 \pi^2} E_2(\si_o/\La^2), ~~~
    T_2 = -{\bra \vsi^2 \ket \over 32 \pi^2} E_1(\si_o/\La^2) + E_1(M^2/\La^2).
    \eeq
For $n \geq 0$, $E_n(z) = \int_1^\infty ~ dt ~ t^{-n}~ e^{-zt} $ \cite{abramowitz-stegun}. Moreover, $E_1(z) \to -\log z - \gamma$, $E_2(z) \to 1 + (\log{z} + \gamma -1)z$ and $E_3(z) \to \half -z + ({3 \over 4} - {\gamma \over 2} - {\log{z} \over 2})z^2$ as $z \to 0$. $T_{0,1,2}$ have leading quartic, quadratic and log divergences as $\La \to \infty$. The divergent, scale-violating and finite terms are listed here, while omitting terms that vanish as $\La \to \infty$:
    \beq
    T_0 &\to& -{\La^4 \over 32 \pi^2} + {\si_o \La^2 \over 16 \pi^2}
    	- {\si_o^2 \over 16 \pi^2 } \bigg(- \half \log{\sigma_o \over \La^2} 
     	+ {3 \over 4} - {\gamma \over 2} \bigg), \cr
    T_1 &\to& {\bra \vsi \ket \La^2 \over 16 \pi^2} 
    	+  {\bra \vsi \ket \si_o \over 16 \pi^2} \bigg( \log(\si_o/\La^2) 
		+ \gamma -1 \bigg), \cr
    T_2 &\to& {\bra \vsi^2 \ket \over 32 \pi^2} \bigg( \log(\si_o/\La^2) 
    	+ \gamma \bigg) - \log(M^2/\La^2) - \gamma.
    \eeq
Thus we isolated the divergent and scale-dependent parts of $\tr \log[-\grad^2 + \si(x)]$ and found they depend locally on $\si(x)$, indeed only on $\vsi(x) = \si(x) - \si_o$ and not on derivatives of $\si(x)$. $W_0(\sigma)$ is chosen to cancel these, so in this regularization scheme $W_0$ is local in $\si(x)$:
    \beq
    W_0(\sigma(x),\La) =  m^2 \si(x) - \si(x)^2/\la 
    	- \hbar (T_0 + T_1 + T_2).
    \eeq
However, the physical consequences of this locality property remain to be studied.

\section{Small oscillations around constant vacua}
\label{s-small-osc}

\subsection{Constant extrema of $N=\infty$ effective action}
\label{s-const-extrema-of-eff-action}

Field configurations extremizing $\Gamma_0(B,\Si)$ (\ref{e-eff-ac-large-N-finite-and-scale-invariant}) dominate the path integral over $b$ and $\si$ in the saddle point approximation. Extrema satisfy the `classical' (large-$N$) equations of motion (eom)
    \beq
    {\delta \Gamma_0}/{\delta B} &=& \{ -\grad^2 + \Si(x) \} B(x) = 0,
    	 \cr
    {\delta \Gamma_0}/{\delta \vsi} &=& {B^2 + m^2 \over 2}
        - {\Si \over \la} - { \hbar \Pi(\Delta) \over {16\pi^2} } \vsi 
        - {\hbar \over 32 \pi^2} (\vsi - \Si \log{\Si/\Si_o}) + \cdots = 0,			\cr
    \dd{\Gamma_0}{\Si_o} &=&  \int d^4x \bigg[ {B^2 + m^2 \over 2}
    	- {\Si \over \la} - {\hbar \over 32\pi^2} \bigg( \vsi 
		\bigg\{\Pi'(\D) {\grad^2 \over \Si_o^2} - \Pi'({\vsi/\Si_o}) {\vsi 			\over \Si_o^2} \bigg\} \vsi + \cdots \bigg) \bigg] = 0.
    \label{e-class-eqn-motion}
    \eeq
The $\cdots$ denote variations of cubic and higher order terms in $\vsi$ that also involve gradients. Roughly, the first two eom determine $B(x)$ and $\vsi(x)$ while the third is needed to fix the constant $\Si_o$, to get the extremal $\Si(x) = \Si_o + \vsi(x)$. Here $\D = -\grad^2/\Si_o$ and
	\beq
	\Pi'(y) = {2(y+1) \log{(y+1)} - y(y+2) \over 2y^3} = -\half - (y+1)(\log(y+1) + {3 \over 2}) + {\cal O}(y+1)^2
	\eeq
Let us begin by looking for constant extrema $B = B_c$ and $\Si = \Si_c$ of $\Gamma_0$, where `c' stands for `constant classical'. The eom become exact since the remaining terms in (\ref{e-class-eqn-motion}) involve gradients:
    \beq
    \Si_c B_c = 0, &&
    \half (B_c^2 + m^2) - {\Si_c \over \la}- {\hbar \over 32\pi^2} 
    	\{ \Si_c - \Si_o - \Si_c \log{(\Si_c / \Si_o)} \} = 0, \cr
	{\rm and} && {B_c^2 + m^2 \over 2} - {\Si_c \over \la} 
		+ {\hbar \over 32\pi^2} 
		{(\Si_c - \Si_o)^3 \over \Si_o^2} 
		\Pi'({(\Si_c - \Si_o )/ \Si_o}) = 0.
    \label{e-class-eqn-motion-constant-flds}
    \eeq
The simplest constant extremum is $\Si_c =0, B_c=0$, with $\Si_o=0$; it exists only if $m=0$ or $\la =0$. This solution is $O(N+1)$ symmetric. Aside from this case, there are two types of extrema based on which factor is non-zero in the $1^{\rm st}$ eom $\Si_c B_c =0$. In the symmetric phase $O(N+1)$ is unbroken and $B_c =0$ and $\Si_c \ne 0$. Solving the eom, the extremum is at $B_c=0, \Si_c = \la m^2/2$ with $\Si_o = \Si_c$ ($\implies \vsi =0$). To attain the symmetric phase, $m^2$ and $\la$ must have the same sign so that the vev $\Si_c$ is non-negative, as $\Si$ is valued on a contour that misses ${\bf R}^{-}$. The extremum at $\Si_c=B_c=0$ is a limiting case when either $m$ or $\la$ vanish.

The broken phase occurs when $\Si_c =0$ and $B_c \ne 0$, where $O(N+1)$ is spontaneously broken to $O(N)$. In this case the $2^{\rm nd}$ and $3^{\rm rd}$ eom can be satisfied only if $\Si_o = \Si_c =0$ (this can be regarded as a limiting case where $\vsi = \Si_c - \Si_o = - \Si_o \to 0$) and $B_c^2 = - m^2$. So the extrema occur at the pair of configurations $B_c = \pm |m|, \Si_c =0$. To realize the broken phase, we need $m^2 < 0$, since the vev $B_c$ of the real field $b$ must be real. This is of course the negative mass$^2$ needed for spontaneous symmetry breaking.

\subsection{Masses of long wavelength small oscillations in symmetric phase}
\label{s-mass-of-lightest-particles-unbroken-phase}

We are interested in small amplitude oscillations around the symmetric phase $\Si_c = \la m^2/2, B_c = 0$. $m^2$ and $\la$ must have the same sign since $\Si$ is valued on a contour that misses ${\bf R}^{-}$. We set $\Si = \Si_c + \gd \si, B = B_c + \gd b$ where $\gd b$ and $\gd \si$ are small compared to the scale of $\Si_c$. Linearizing the eom (\ref{e-class-eqn-motion}) we get
	\beq	
	-\grad^2 \gd b + (\la m^2 / 2) \gd b =0, ~~~
	- \hbar (16 \pi^2)^{-1} {\Pi(\D)} \gd \si - \gd \si/\la =0 {\rm ~~and~~}
	\int d^4x ~\gd \si =0.
	\eeq
If we further restrict to the longest wavelength oscillations, then $\gd \si$ is slowly varying compared to $\Si_c$. Ignoring fourth and higher derivatives, we get
	\beq
	-\grad^2 \gd b + (\la m^2 / 2) \gd b =0, ~~
	-\grad^2 \gd \si - \hbar^{-1} 48 \pi^2 m^2~ \gd \si =0 {\rm ~~and~~} 
	\int d^4x ~\gd \si =0.
	\eeq
Wick rotating back to Minkowski space $t = -i \tau$, these are a pair of Klein-Gordon equations $(\pdr_t^2 - \pdr_{\bf x}^2 + M_{b,\si}^2)(\gd b, \gd \si) =0$ for particles of masses $M_b^2 = \la m^2/2$ and $M_\si^2 = -48\pi^2 m^2/\hbar$. There could also be heavier particles corresponding to shorter wavelength oscillations of $\delta \sigma$. The third eom $\int d^4x ~\gd \si =0$ implies that the constant part of $\gd \si$ vanishes, it is satisfied by the wave-like solutions of the Klein-Gordon equation. The oscillations of $b$ are linearly stable provided $m^2$ and $\la$ have the same sign, while those of $\sigma$ are linearly stable for $m^2 < 0$ and unstable for $m^2 > 0$. In other words, for $\hbar >0$, the symmetric phase is linearly stable to long-wavelength small perturbations provided $m^2, \la < 0$ (the boundary of this region has neutral stability).

When $\hbar \to 0$, the oscillations of $\gd \sigma$ are expelled from the spectrum and we only have the $b$ particle of mass $M_b^2 = \la m^2/2$. This is to be expected from the potential $V(\eta) = {\la \eta^2 / 4} + \la m^2 \eta / 2$ corresponding to $W(\si) = m^2 \si - \si^2/\la$. The $\gd b$ oscillations are linearly stable as long as $\la$ and $m^2$ have the same sign. If $\la , m^2 < 0$ these oscillations are non-linearly unstable in the $\hbar \to 0$ limit. They are just the metastable oscillations around the symmetric minimum of an `M' shaped potential, which is not bounded below. On the other hand, if $\la, m^2 > 0$, the symmetric phase is a global minimum and the $\gd b$ oscillations are absolutely stable. A more careful non-linear stability analysis is necessary for $\hbar > 0$.

\subsection{Long wavelength small oscillations in the broken phase}
\label{s-broken-phase}

Here we wish to study the longest wavelength small oscillations around the symmetry broken vacua $\Si_c =0, B_c^2 = - m^2$. For small oscillations, we write $B = B_c + \gd b, \Si = \gd \si = \gd \Si_o + \gd \vsi$, where $\gd b$ and $\gd \si$\footnote{$\Si$ is valued on a contour that misses ${\bf R}^-$, so $\gd \si$ does not take negative values, $|\gd \si| \ll B_c^2$. $\gd \Si_o$ is a positive constant, which is not necessarily small compared to $B_c^2$ and heuristically $\gd \Si_o \sim {\cal O}(B_c^2)$. $\gd \vsi$ could also have a constant part, which could be negative enough to cancel $\gd \Si_o$. These must be determined by solving the eom. For slowly varying perturbations we drop quadratic and higher terms in $\grad^2 / \gd \Si_o$.} are small and slowly varying on the scale set by $B_c$. The eom. become:
	\beq
	-\grad^2 \gd b + B_c \gd \si =0, &&
	B_c \gd b - \bigg[ \ov{\la} + {\hbar \over 16\pi^2} \Pi({-\grad^2 \over \gd \Si_o}) \bigg] 
		\gd \sigma - {\hbar \over 32 \pi^2} \bigg[\gd \vsi 
		- \gd \si \log(\gd \si / \gd \Si_o) \bigg] = 0 
	\cr	{\rm and}~~~ 
	\int d^4x && \bigg[B_c \gd b - {\gd \si \over \la} - {\hbar \over 32 \pi^2} \bigg\{
		(\gd \vsi)~  \Pi'({-\grad^2 \over \gd \Si_o}) {\grad^2 \gd \vsi \over \gd \Si_o^2} 
		- \Pi'({\gd \vsi \over \gd \Si_o}) {\gd \vsi^3 \over \gd \Si_o^2 } \bigg\} \bigg] = 0.
	\label{e-eom-linearized-broken-phase}
	\eeq
We can eliminate $\gd b$ by taking the laplacian of the 2nd eom (using $\grad^2 \gd b = B_c \gd \si$):
	\beq
	-\bigg[ \ov{\la} -  {\hbar \over 32 \pi^2} \log{\gd \si \over \gd \Si_o} 
		+ {\hbar \over 16\pi^2} \Pi({-\grad^2 \over \gd \Si_o}) \bigg] \grad^2 \gd \si  
		+ {\hbar \over 32 \pi^2} {|\grad \gd \si|^2 \over \gd \si} 
		- m^2 \gd \si  =0.
	\eeq
If  we restrict to long wavelength oscillations and keep only two derivatives, we get
	\beq
 	-\bigg( \ov{\la} -{\hbar \over 32 \pi^2} 
		\log{\gd \si \over \gd \Si_o} \bigg) \grad^2 \gd \si 
		+ {\hbar \over 32 \pi^2} {|\grad \gd \si|^2 \over \gd \si} 
		- m^2 \gd \si = 0.
	\label{e-non-lin-eqn-osc-broken-phase}
	\eeq
Here $\gd \si$ is valued on a contour that misses ${\bf R}^{-}$ and $\gd \Si_o > 0$, so the $\log$ and quotients make sense. In deriving (\ref{e-non-lin-eqn-osc-broken-phase}), we ignored the ${\cal O}(\vsi^3,\grad^2)$ and higher order terms in $\Gamma_0$ (\ref{e-eff-ac-large-N-finite-and-scale-invariant}) which contain derivatives. They would contribute terms with $>2$ derivatives, just as $\Pi(-\grad^2/ \gd \Si_o)$, since we took the Laplacian of the 2nd eom. But we could not ignore the $\Pi(\vsi/\Si_o)$ term in $\Gamma_0$, indeed it is responsible for all the non-linearities and $\hbar$-dependence. We must solve (\ref{e-non-lin-eqn-osc-broken-phase}) for $\gd \si$ and recover $\gd b$ using using $\grad^2 \gd b = -m^2 \gd \si$ and self-consistently fix the constant $\gd \Si_o$ using the 3rd eom in (\ref{e-eom-linearized-broken-phase}). Though we study small oscillations, the non-linear terms may not be negligible, which reminds us of the KdV equation. We hope to study this challenging problem elsewhere and only consider the simplest case here. When $\hbar =0$, the 2nd eom reduces to $B_c \gd b - \gd \si/\la =0$ which implies the 3rd eom. So in this case $-\grad^2 \gd b - m^2 \la \gd b =0$, corresponding to Klein-Gordon oscillations of a (Higgs) particle of mass $M_H^2 = -m^2 \la$. For linear stability $\la m^2 <0$ and $m^2 < 0$ in the broken phase, so $\la > 0$. These are oscillations around the symmetry-broken minima of the Mexican hat potential $V(\eta) = {\la \eta^2 /4} + \half \la m^2 \eta$ with $m^2 < 0$ and $\la > 0$.

\section{Large-$N$ fixed-points in lower dimensions}
\label{s-2d}

The principle of scale-invariance may also be implemented in $d=2$, where $S = \ov{2\hbar} [\hbar \tr \log{-\grad^2 + \sigma \over M^2} + \int d^2x \{ (\grad b)^2  + \si b^2 + W(\sigma) \} ] $ as before. As $N \to \infty$, in $\zeta$-regularization, the choice $\int d^2x W_0(\Si) = \hbar \Om \Si_o (4\pi)^{-1} \log(\Si_o/M^2) (1 + {\bra \vsi \ket \over \Si_o}) -\la \Om \bra \Si \ket$ makes the effective action finite and scale-invariant for a dimensionless coupling $\la$, with $\Om = \int d^2x$. The corresponding effective action is (here $\D = -\grad^2/\Si_o$, $\Pi_2(\D) = \D \Pi^\pr(\D) = \{2(1+\D) \log(1+\D) -\D(2+\D) \}/2\D^2$ and $\Si = \Si_o + \vsi$)
	\beq
	{2 \Gamma_0 \over \Om} = \bra (\grad B)^2 + \Si B^2 - \la \Si 
		+ {\hbar \Si_o \over 4\pi} \{1 - {\vsi^2 \over 2 \Si_o^2} \}
	- {\hbar \over 4\pi \Si_o} \bigg\{ \vsi (\Pi_2(\D) + \Pi_2({\vsi \over \Si_o} )) \vsi 
	+ {\cal O}(\vsi^3, \grad^2) \bigg\} \ket 
	\eeq
The constant extrema of $\Gamma_0$ are: (S) if $\la = \hbar/4\pi$, then $b_c = 0$, $\sigma_c$ is arbitrary and $O(N+1)$ is unbroken; (B) if $\la > \hbar/4\pi$, then $\si_c =0$, $b_c = \pm \sqrt{\la-\hbar/4\pi}$ and $O(N+1)$ is spontaneously broken. Upon inverting the Laplace transform at $N = \infty$, the $\zeta$-regularized part of the interactions $W_0(\si) = {\hbar \si \over 4\pi} \log({\si \over M^2}) - \la \si$ correspond to the family of potentials $V_0(\eta) = -{\hbar M^2 \over 4\pi} \exp{[-{4\pi (\eta - \la) \over \hbar}-1]}$ where $\eta = {|\phi|^2 \over N}$. As in $d=4$, we have a line of quantum fixed-points at large-$N$. When $\hbar \to 0$ these reduce to the massless free theory, the only classical fixed-point. 



Odd $d=3$ is different, as the scale anomaly from quantum fluctuations vanishes. Indeed,
$\zeta_{-\grad^2 + \sigma}(0) =0$ and there is no scale-dependence$(M)$ in the $\zeta$-regularized finite part of 
	\beq
	\tr \log[{-\grad^2 + \sigma \over M^2}] = - {\Om \si_o^{3/2} \over
		(4\pi)^{3/2}} \sum_{n=0}^\infty {\bra a_n \ket \over \si_o^n}
		\Gamma(n-{3 \over 2}) = - {\Om \si_o^{3/2} \over
		(4\pi)^{3/2}}\bigg[{4 \sqrt{\pi} \over 3} 
		+ {2 \sqrt{\pi} \bra \vsi \ket \over \si_o} 
		+ {\sqrt{\pi} \bra \vsi^2 \ket \over 2 \si_o^2} + \cdots \bigg]
	\eeq 
With no scale anomaly to cancel, a scale-free $\Gamma_0$ results if $W_0(\sigma) = - \si^{3/2}/\la$, with dimensionless $\la$. The finite part of the original potential corresponding to $W_0$ is $V(\eta) = {4 \la^2 \eta^3 \over 27}$, i.e. the large-$N$ limit of the $|\phi|^6$ theory\footnote{Unlike in $d=2,4$ where $W_0(\si)$ included scale-violating terms, in $d=3$ it is scale-invariant and a first approximation to $\Gamma_0$, so $V(\eta)$ is a good way of specifying the theory.}. So our principle applied in $d=3$ implies a line of quantum mechanical large-$N$ fixed-points corresponding to the $|\phi|^6$ interaction\footnote{There is a difference between even and odd $d$. The line of fixed-points in $d=3$ includes the GFP as a special case ($\la =0$) for any $\hbar$. In $d=2,4$ the GFP lay on the line of fixed-points only when $\hbar =0$.}. This agrees with perturbative results that $\la$ is exactly marginal in the large-$N$ limit ($\beta$ function vanishes) \cite{Townsend-phi-sixth,phi-sixth-3d}. However, perturbatively, only a pair of these fixed-points survive the first $1/N$ corrections, the trivial one and a non-trivial UV fixed-point. A non-perturbative analysis\cite{BMB} modifies this picture, but suggests the existence of the Bardeen-Moshe-Bander large-$N$ UV fixed-point, see \cite{phi-sixth-3d}.


\section{Discussion and open problems}
\label{s-discussion}

We argued that a model constructed as a mass deformation of a non-trivial fixed-point would solve both UV and naturalness problems of $4d$ $O(N+1)$ scalar fields. Moreover, unlike breaking SUSY (which may produce new naturalness problems, \rlap{ / }{CP} phases), breaking scale-invariance by a mass term is harmless, as is breaking chiral symmetry by an electron mass in QED. At $N=\infty$, we found a line of non-trivial fixed-points with finite and scale-invariant effective actions $\Gamma_0$ parameterized by a coupling $\lambda$. They reduce to scale-invariant classical $\lambda \phi^4$ theory when $\hbar \to 0$. The model isn't built via small quantum corrections to a pre-existing classical theory, since `action' and `quantum fluctuations' are comparable. For $\hbar > 0$, the potential $V$ leading to $\Gamma_0$ doesn't approximate the effective potential; its minima have no physical meaning. Unlike $\Gamma_0$, neither $V$ nor `$[D\phi]$' is finite without regulators. They also depend on a scale $M$, which mutually cancels. At large-$N$, the {\em finite part} of $V$ in $\zeta$-regularization grows as $V(\fr{|\phi|^2}{N}) \sim \fr{|\phi|^4/N^2}{\log(|\phi|^2 \bar \la(\hbar) / M^2 N)}$ as $|\phi|^2/N \to \infty$ (appendix \ref{a-back-to-V-legendre-transform}). $V(\eta)$ is best expressed in terms of the Laplace transformed potential $W_0(\si)$ (\ref{e-W_0}). In sec. \ref{s-locality-of-W}, we showed that all terms in $W_0(\si)$ that cancel divergences and scale violations from quantum fluctuations are local. $\Gamma_0$ (\ref{e-eff-ac-large-N-finite-and-scale-invariant}) was found in an expansion around a constant background field. Since $\Gamma_0$ incorporated {\em all} quantum fluctuations of $\phi_{1, \ldots, N}$, it involved vertices of all orders. $\Gamma_0$ was scheme dependent, to relate two schemes for constant backgrounds, $1/\la$ is shifted by a finite additive constant (sec. \ref{s-mom-cut-off-regularization}, appendix \ref{a-dim-reg}). Extremizing $\Gamma_0$ after adding a mass deformation revealed vacua where $O(N+1)$ is unbroken or spontaneously broken to $O(N)$. We calculated masses of lightest excitations and derived an intriguing non-linear equation (\ref{e-non-lin-eqn-osc-broken-phase}) for oscillations about the broken phase. Masses could be naturally small due to dilation invariance when they vanish. Roughly, our fixed-points lie on a plane parallel to and a distance $\propto \hbar$ from the $m-\la$ plane of $m^2 \phi^2 + \la \phi^4$ theory.

In $3$d, our construction reduced to a known result that $|\phi|^6$ is scale-invariant at large-$N$, giving us confidence to apply it in $d=2,4$. To get a scale-invariant $\Gamma$ we canceled scale anomalies from quantum fluctuations by choosing an $\hbar$-dependent action. We do not advocate unrestricted choice of action to cancel any divergences. Rather, it is determined by the principle of scale-invariance and $W_0(\si)$ is not fine-tuned. Why don't we usually cancel anomalies from quantum fluctuations by a choice of action? It is a {\em physical} question. If we aim to model a system displaying a symmetry despite the presence of potential anomalies from quantum fluctuations (as we have argued a light scalar would indicate), then it is desirable to cancel them. Elsewhere, if we wish to model a quantum system exhibiting a symmetry violation (eg. $\pi^0 \to 2\gamma$, chiral symmetry), then anomalies from quantum fluctuations must not be canceled. Furthermore, actions depending on $\hbar$ are not new. In SUSY quantum mechanics $H = \half (p^2 + W^2(x) + \hbar \si_3 \dd{W}{x})$ includes a `Yukawa' interaction $\propto \hbar$, crucial for SUSY and cancelation of vacuum energy \cite{witten-dyn-susy-brk}. In quantum Liouville theory, correlations are expected to exhibit a $b \to (\hbar b)^{-1}$ symmetry and 2d lattice of poles, based on a conjectured solution of the conformal bootstrap equations\cite{dorn-otto-zamolodchikov}. However, classical Liouville theory (potential $\mu_b e^{2 b \phi}$), when quantized by path integrals, doesn't exhibit this symmetry. But by postulating an $\hbar$-dependent potential $\mu_b e^{2 b\phi} + \mu_{(\hbar b)^{-1}} e^{2 \phi / \hbar b}$, the conjecture was proved with the symmetry and lattice of poles \cite{liam-sreedhar}.

We mention some open problems now. (1) We would like to know whether any of our fixed-points survives at finite $N$ and whether the model retains predictive power at higher orders in $1/N$. The situation can be quite subtle, as investigations of $3d$ $|\phi|^6$ theory indicated \cite{phi-sixth-3d}. (2) A study of the non-linear equation we derived for oscillations around the broken phase is needed, along with a better understanding of the contour on which $\sigma$ is valued. (3) We used the naive scaling dimensions of $\phi,b,\si$ to define scale-invariance of the effective action. This was the simplest physical possibility and may be a good approximation for large-$N$ and small $\hbar$. But in general we must allow for anomalous dimensions. (4) A trivial theory can look non-trivial, so a more careful investigation of our fixed-points is needed. We must compute correlation functions, dimensions of composite operators and the effects of $1/N$ corrections. (5) It is interesting to couple our scalars to fermions (especially the top quark which has the largest Yukawa coupling). Even if $1/N$ corrections eliminate the UV fixed-point in the scalar sector, there could be one after including fermions/gauge fields. (6) We wonder whether there is a dual description of our scale-invariant model by analogy with  AdS/CFT \cite{string-dual}. (7) Implications of the possible large-$N$ fixed-points in $d=2$ remain to be studied. (8) Since symmetry breaking is well-described by $\la \phi^4$ at low energies, it may be phenomenologically interesting to build a model governed by a cross-over from the trivial fixed-point to a non-trivial fixed-point of our sort. (9) It would be useful to find some regime where a form of perturbation theory can be used to study our model, perhaps for small $\hbar$ and $\lambda$. (10) Does the presence of scaling symmetry at $m=0$ protect $m$ from large ($1/N$) corrections? (11) The functional RGE may provide a complementary way to test our proposal. (12) A numerical search for our large-$N$ fixed-points would also be interesting.

\section*{Acknowledgements}

This work was supported by a UK-EPSRC fellowship and earlier by a Marie Curie fellowship.

\appendix

\section{Examples of naturalness}
\label{a-eg-naturalness}

By a naturalness explanation for a small quantity, we mean that the model acquires an additional symmetry when that quantity vanishes \cite{tHooft-naturalness}. In the absence of such a symmetry, its natural value is $\sim 1$ in units of the microscopic scale where the model is superseded. The symmetry may be continuous/discrete. The actual small value of the quantity (if $\ne 0$) is not predicted by this principle and usually requires a microscopic theory for its determination. But its effects can often be treated perturbatively, for example by the introduction of symmetry breaking terms in the action. Below are some examples of naturalness explanations, it appears this concept explains several small parameters both in tested theories and mathematical models\footnote{In \cite{richter}, Richter criticized naturalness. However, the definition via symmetries is not the one he uses. His criticisms seem to have more to do with the large number of parameters in the supersymmetric standard model.}. Indeed, besides $m_H$, there is a naturalness explanation for most small parameters in the standard model. This gives us confidence to turn things around: if there is an unreasonably small parameter in nature or in a model, then there must be some symmetry, which if exact, would make that parameter vanish. Thus, naturalness can be useful in model building\footnote{Besides approximate symmetries, there could be other mechanisms (tunneling) responsible for small quantities.}.

{\bf (1)} The fact that planetary orbits are nearly closed and nearly lie on a plane, are related to the conservation of angular momentum and Laplace-Runge-Lenz vector in the Kepler problem.
{\bf (2)} Near-degeneracies of energy levels in atomic spectra: In hydrogen-like atoms, $E_{nlm} - E_{nlm'} =0$ due to spherical symmetry. Small energy difference can be due to direction of magnetic field breaking spherical symmetry.
{\bf (3)} In hydrogen-like atoms, the `accidental degeneracy' of energy levels with the same value of $l$ is due to a hidden $SO(4)$ symmetry whose conserved quantities are angular momenta and Laplace-Runge-Lenz vectors.
{\bf (4)} Some near-degeneracies in atomic energy levels can be explained by parity invariance of electrodynamics. The small splittings are due to parity violation in the weak interactions \cite{atomic-parity-viol}.
{\bf (5)} The imaginary parts of eigenvalues of several non-hermitian Schrodinger operators vanish due to an unbroken PT symmetry \cite{bender-PT}.
{\bf (6)} Near degeneracies $m_n - m_p \simeq 1.29$ MeV and $m_{\pi^{\pm}} - m_{\pi^0} \simeq 4.59$ MeV: if isospin were an exact symmetry, $n,p$ would be degenerate in mass (as would $\pi^{\pm,0}$). Isospin breaking by quark mass difference and electromagnetic interactions explain the small $n$-$p$ and $\pi^\pm$-$\pi^0$ splittings.
{\bf (7)} Pions are naturally light compared to $\rho$ mesons due to chiral symmetry. Pions are pseudo-goldstone bosons of spontaneously broken chiral symmetry. If the quarks were massless, chiral symmetry would be exact at the level of the lagrangian, and be spontaneously broken to ${SU(N_f)}_V$, and the pions would be massless goldstone bosons. But non-zero current quark masses explicitly break chiral symmetry and give the pions a small mass calculable via chiral perturbation theory.
{\bf (8)} Experimentally, the mass of a photon is less than $10^{-16}$ eV outside a superconductor\cite{pdg}. This is explained by the exact $U(1)$ gauge symmetry if the photon is massless.
{\bf (9)} Small $m_e$: If $m_e =0$, QED gains chiral symmetry. Same applies to $m_\mu, m_\tau$, there is a different chiral symmetry for each. Smallness of $m_e / m_\tau$ remains unexplained. 
{\bf (10)} Small current quark masses: For $N_{f} \geq 2$, if current quark masses $\to 0$, QCD gains a partial unbroken chiral symmetry $SU(N_f)_{V}$.
{\bf (11)} Small neutrino masses: Chiral symmetry for each flavor is exact if neutrinos are massless.
{\bf (12)} Parity is an exact symmetry of QCD in the absence of the topological $\theta$ term, which is parity odd. Thus, a small QCD $\theta$-angle is natural within the theory of strong interactions.
{\bf (13)} Radiative corrections to $m_W/m_Z$ are small. If they were zero, the standard model would have custodial symmetry. The small radiative corrections come from the gauge interactions which do not respect the custodial symmetry ($O(4)$ symmetry of the scalars, spontaneously broken to $O(3)$). Naively, one expects $m_W/m_Z$ to receive radiative corrections in the scalar self coupling $\la \approx {2 m_H^2 /(246 {\rm GeV})^2}$, which could be large. But custodial $O(3)$ symmetry forces these to vanish.
{\bf (14)} Small coupling constants can be explained by the separate conservation laws for particles, gained by setting their coupling to zero.
{\bf (15)} The inverse of correlation length is very small near a continuous phase transition. This is natural due to the emergence of scaling symmetry.
{\bf (16)} Some linear combinations of correlations in large-$N$ multi-matrix models vanish because of the presence of hidden non-anomalous symmetries \cite{ward-id-sde}.
{\bf (17)} $m_{\rm Higgs}$ in a SUSY standard model\cite{SUSY-higgs} can be naturally small. If $m_H =0$, we have unbroken global SUSY (when the super-partner fermion is also massless, which is natural by chiral symmetry).
{\bf (18)} It was suggested in\cite{tHooft-nobbenhuis} that a discrete symmetry relating real to imaginary space-time coordinates could ensure a naturally small cosmological constant.

\section{Large-$N$ effective potential via dimensional regularization}
\label{a-dim-reg}

To calculate $\tr \log [-\grad^2 + \Si_o] = \Om \int [d^4p] \log[p^2 + \Si_o]$ appearing in $\Gamma_0$ (\ref{e-def-of-eff-ac-at-infinite-N}), continue to $n$ dimensions and differentiate, to get a convergent integral for $n < 2$
    \beq
    T_n &=& \int {d^n p \over (2\pi)^n} \log[p^2 + \Si_o] ~~\implies~~
    \dd{T_n}{\Si_o} = \int {{d^n p / (2\pi)^n} \over (p^2 + \Si_o)} =
    (4\pi)^{-n/2}~ {\G(1-n/2) \over \Si_o^{1-n/2}}
	\cr
    \dd{T_n}{\Si_o} &=& {\Si_o \over 8 \pi^2 (n-4)} + {(\g -1 + \log[\Si_o/4\pi])	\Si_o \over 16 \pi^2} + {\cal O}(n-4).
    \eeq
Now $\hbar \tr \log[-\grad^2 + \Si_o] = \hbar \Om T_n$. So integrating with respect to $\Si_o$,
    \beq
    \tr \log[-\grad^2 + \Si_o] = {\Om \Si_o^2 \over 16 \pi^2
    (n-4)} + {\Om \Si_o^2 \over 32 \pi^2}(\g - {3 \over 2} - \log
    4\pi) + {\Om \Si_o^2 \log \Si_o \over 32 \pi^2} + c \Om + {\cal O}(n-4).
    \eeq
$c$ (independent of $\Si_o$) only adds a constant to the effective potential. We have a pole part, finite part and terms that vanish as $n \to 4$. The finite part that transforms inhomogeneously under rescaling ${\hbar \Om \Si_o^2 \log \Si_o \over 32 \pi^2}$ is the same as in cutoff or $\zeta$-regularization. The choice of $W_0$ that makes $\Gamma_0$ finite and scale-free for any $\la$ in the limit $n \to 4$ is
    \beq
    W_0(\Si_o,n) &=& -{\hbar \Si_o^2 \over 16 \pi^2 (n-4)} 
    	- {\hbar \Si_o^2 \log \Si_o \over 32 \pi^2} - \Si_o^2 / \la \cr
    {\rm and ~~~} \G_0(B_o,\Si_o) &=& (\Om / 2) \bigg[\bigg(- \ov{\la}
        + {\hbar (\g - 3/2 - \log 4\pi) \over 32 \pi^2} \bigg) \Si_o^2
        + \Si_o B_o^2 \bigg].
    \eeq
$\la$ however is scheme dependent $\la_\zeta^{-1} = \la_{\rm dim-reg}^{-1} + {\hbar (3/2 + \log 4\pi - \g)}/32 \pi^2.$

\section{Expansion of $\tr \log[-\grad^2 + \si]$ in $\vsi(x) = \si(x) - \si_o$}

\label{a-exp-for-tr-log}

\subsection{Zeta function in terms of the heat kernel}
\label{a-zeta-fn-heat-kernel}

Let $A = -\grad^2 + \si(x)$ and $\zeta_A(s) = \tr A^{-s}$. Then $\tr \log A = -\zeta'(0)$. We get an integral representation for $\zeta_A(s)$ by making a change of variable $t \mapsto At$ in the formula for $\Gamma(s)$:
    \beq
        A^{-s} = {\Gamma(s)}^{-1} \int_0^\infty dt ~e^{-tA}~ t^{s-1}.
    \eeq
If $B= {(-\grad^2 + \si) / M^2}$, then $\tr \log B = \tr \log A - \zeta(0) \log{M^2}$ where $\tr \hat 1 := \zeta(0)$ is calculated in \ref{a-scale-anomaly}. Now define the evolution operator $\hat h_t = e^{-tA}$ which satisfies a generalized heat equation
    \beq
        \pdr_t~ \hat h_t = -A \hat h_t = ( \grad^2 - \si) ~\hat h_t, 
        ~~~~ \hat h_{t \to 0^+} = \hat 1
    \eeq
It is convenient to work with the heat kernel $\hat h_t ~\psi(x) = \int ~d^d y~ h_t(x,y) ~\psi(y)$ which satisfies
    \beq
        \pdr_t~ h_t(x,y) = [~\grad^2 - \si(x)~] ~h_t(x,y) {~~~\rm and ~~~}
         h_{t\to 0^+}(x,y) = \delta^d(x-y).
    \label{e-gen-heat-eqn}
    \eeq
Then $\zeta_A(s)$ is the Mellin transform of the trace of the heat kernel:
    \beq
        \zeta_A(s) = \tr A^{-s} = {\Gamma(s)}^{-1} 
        	\int_0^\infty dt ~t^{s-1} ~\tr e^{-tA}
        = {\Gamma(s)}^{-1} \int_0^\infty dt ~t^{s-1} ~ \int d^d x ~h_t(x,x).
    \label{e-def-of-zeta-as-mellin-transf}
    \eeq
To find $h_t(x,x)$ we need to solve (\ref{e-gen-heat-eqn}). For constant complex $\si = \si_o$, (\ref{e-gen-heat-eqn}) the solution is $h^o_t(x,y) =(4 \pi t)^{-d/2}  \exp{\{-t \si_o -{(x-y)^2/ 4t}\}}.$


\subsection{Short time expansion for heat kernel}
\label{s-short-time-exp-heat-kernel}

Now we expand $h_t(x,y)$ in derivatives and powers of $\vsi(x) = \si(x) - \si_o$ for small $t$ \cite{heat-kernel-expansion}. Assuming that its `non-analytic part' is captured by the case $\si = \si_o$ we make the ansatz
    \beq
        h_t(x,y) = h_t^o(x,y) ~\sum_{n=0}^\infty a_n(x,y) t^n  =
        {e^{-\si_o t} e^{-(x-y)^2/4t}} (4\pi t)^{-d/2} ~\sum_{n=0}^\infty a_n(x,y)
        t^n.
    \label{e-ansatz-for-heat-kernel}
    \eeq
The average value of $\vsi$ need not vanish. But, we assume that $\grad \vsi(x) \to 0$ as $|x| \to \infty$ so that $\int d^dx ~(\grad^2)^p \vsi^q(x) = 0$ for $p,q \geq 1$. For (\ref{e-ansatz-for-heat-kernel}) to satisfy initial condition (\ref{e-gen-heat-eqn}), $a_0 = 1$. If $\si$ is a constant, $a_i = \delta_{0,i}$. The $e^{-\si_o t}$ in (\ref{e-ansatz-for-heat-kernel}) makes the Mellin transform (\ref{e-def-of-zeta-as-mellin-transf}) convergent for $\Re \si_o > 0$, which is necessary to recover $\zeta_A(s)$. Putting (\ref{e-ansatz-for-heat-kernel}) into (\ref{e-gen-heat-eqn}) gives
    \beq
        \sum_0^\infty (n+1) ~a_{n+1} ~t^n =
        -(x-y)_i ~\sum_0^\infty ~t^n \grad_i ~a_{n+1}
        + \sum_0^\infty ~ t^n \grad^2 a_n
        - \vsi \sum_0^\infty a_n t^n.
    \eeq
Comparing coefficients of $t^n$ determines $a_{n+1}$ given $a_n$ and the initial condition $a_0 = 1$
    \beq
        \bigg\{(x-y)_i \grad_i + n + 1 \bigg\} a_{n+1}(x,y) = (\grad^2 - \vsi)
        a_n(x,y).
    \eeq
Now only $a_n(x,x)$ appear in $\zeta(s)$, so we specialize to $a_{n+1}(x,x) = \ov{(n + 1)}(\grad^2 - \vsi) a_n(x,x)$. The first few $a_n(x,x)$ are $a_1 = -\vsi(x)$, $a_2 = \half (\grad^2 - \vsi) a_1 = \half (\vsi^2 - \grad^2 \vsi)$,
    \beq
        a_3 &=& (\grad^2 - \vsi) a_2 / 3 = (\vsi \grad^2 \vsi - \vsi^3
            + \grad^2 \vsi^2 - (\grad^2)^2 \vsi)/3! \cr
        a_4 &=& {(\grad^2 - \vsi) a_3 \over 4} = \ov{4!} 
        	(\grad^2(\vsi \grad^2 \vsi) - \grad^2 \vsi^3 + \grad^4 \vsi^2
            - \grad^6 \vsi - \vsi^2 \grad^2 \vsi
            + \vsi^4 - \vsi \grad^2 \vsi^2 + \vsi \grad^4 \vsi).
    \eeq
Recall that $h_t(x,x) = {e^{-\si_o t}} ~(4\pi t)^{-d/2}~ \sum_0^\infty a_n t^n$. But for $\zeta_A(s)$ we only need $\bra a_n \ket = \int d^d x a_n(x,x) / \int d^d x$. Assuming $\vsi \to {\rm const}$ and $\grad \vsi \to 0$ as $|x| \to \infty$,
    \beq
        \bra a_0 \ket = 1, ~~
        \bra a_1 \ket = - \bra \vsi \ket, ~~
        \bra a_2 \ket = \ov{2!} \bra \vsi^2 \ket, ~~
        \bra a_n \ket = \ov{n!} \bra \vsi (\grad^2)^{n-2} \vsi 
        	+ (-1)^n \vsi^n \ket ~~ n = 3,4,5,\ldots
    \label{e-list-of-avg-val-of-a_n}
    \eeq
where {\em we have ignored cubic and higher order terms in $\vsi$ that also carry gradients.}

\subsection{Derivative expansion for $\tr \log[-\grad^2 + \si]$}
\label{a-derivative-exp-of-tr-log}

We use (\ref{e-def-of-zeta-as-mellin-transf}) and the expansion (\ref{e-ansatz-for-heat-kernel}) to get an expansion for $\zeta_A(s)$ in derivatives of $\si = \si_o + \vsi$
    \beq
        \zeta(s) &=& \ov{\Gamma(s)} \int_0^\infty dt
            \int d^d x ~t^{s-1} ~h_t(x,x) =
        \ov{\Gamma(s)} \int d^d x \int_0^\infty dt~ t^{s-1} {e^{-\si_o t}
            \over (4\pi t)^{d/2}} \sum_0^\infty a_n(x,x) t^n \cr
        \zeta(s)  &=& \Om ~({(4\pi)^{d/2} \Gamma(s)})^{-1}~ 
        	\sum_0^\infty \bra a_n \ket
            \int_0^\infty dt~ t^{s+n-1-d/2} e^{-\si_o t}
    \eeq
where $\Om = \int d^d x$. The integral over $t$ is a Gamma function, and we specialize to $d=4$:
    \beq
        {\zeta(s) \over \Om} &=& 
         {\si_o^{d/2-s} \over (4\pi)^{d/2}} 
        	\sum_n {\bra a_n \ket \over \si_o^{n}}
            {\Gamma(s+n-d/2) \over \Gamma(s)} \cr
        &=& {\si_o^{2-s} \over 16\pi^2} \bigg[{\bra a_0 \ket \over
            (s-1)(s-2)} + {\bra a_1 \ket \over (s-1) \si_o} + {\bra a_2 \ket \over 				\si_o^2} 
            + {s \bra a_3 \ket \over \si_o^3} 
            + {s(s+1) \bra a_4 \ket \over \si_o^4} \cdots \bigg].
	\label{e-expansion-for-zeta-of-s}
    \eeq
Differentiating and setting $s=0$ we get
    \beq
    {\zeta^{\prime}(0) \over \Om} &=& -{\si_o^2 \log{\si_o} \over 16 \pi^2} 
    	\bigg[ {\bra a_0 \ket \over 2}  - {\bra a_1 \ket \over \si_o} 
		+ {\bra a_2 \ket \over \si_o^2} \bigg] 
		+ {\si_o^2 \over 16 \pi^2} \bigg[{3\bra a_0 \ket \over 4} 
		- {\bra a_1 \ket \over \si_o} 
		+ \sum_{n=3}^\infty { (n-3)! \bra a_{n} \ket \over \si_o^{n}}
    \bigg] \cr
     &=& -{\si_o^2 \over 16 \pi^2} \bigg[ {\bra a_0 \ket \over 2} 
     	\log[\si_o e^{-3/2}] + {\bra a_1 \ket \over \si_o} (1-\log{\si_o}) 
		+ {\bra a_2 \ket \over \si_o^2} \log{\si_o} 
		- \sum_{n=3}^{\infty} {(n-3)! \bra a_n \ket \over \si_o^{n}} \bigg]
       \eeq
Inserting expressions for $\bra a_n \ket$ from (\ref{e-list-of-avg-val-of-a_n}), we get a formula for $\tr\log[-\grad^2 + \si] = -\zeta^{\prime}(0)$.
	\beq  
	    -\zeta^{\prime}(0) =  {\si_o^2 \Om \over 32\pi^2} \log{\si_o \over
        e^{3/2}} + \int {d^4x \over 16 \pi^2} \bigg[\vsi \si_o \log{\si_o \over e} 
            + {\vsi^2 \over 2} {\log\si_o} 
            -\sum_{n=3}^\infty {\vsi (\grad^2/\si_o)^{n-2} \vsi 
            + \si_o^2 (-\vsi/\si_o)^n \over n(n-1)(n-2) }
            \bigg].
    \eeq
The sum over $n$ can be performed. Let $\Delta = -{\grad^2 / \si_o}$ or $\vsi/\si_o$ as appropriate, then
    \beq
        \Pi(\Delta) = \sum_{n=1}^\infty {(-\Delta)^n
            \over n(n+1)(n+2)} = {\Delta(3 \Delta +2) -2(\Delta+1)^2
            \log{(1+\Delta)} \over 4 \Delta^2}.
    \label{e-inv-sigma-propagator}
    \eeq
$\Pi(\Delta)$ is analytic at $\Delta=0$,~~ $\Pi(\Delta) = -{\Delta \over 6} + {\Delta^2 \over 24} - {\Delta^3 \over 60} + \cdots$. For large $\D$ ($\si_o \notin {\bf R}^{-}$, $\grad^2 < 0$),
    \beq
    \Pi(\D) \to -\half \log \D + {3 \over 4} - {\log \D \over \D}
    + {\cal O}(\D^{-2}).
    \eeq
The final result, using $\tr \log M^2 = \zeta(0) \log M^2$ from \ref{a-scale-anomaly} is ($\si(x) = \si_o + \vsi(x)$),
    \beq
    \tr\log{\si -\grad^2\over M^2} = {\si_o^2 \Om \over 32\pi^2}
        \log{\si_o e^{-{3 \over 2}} \over  M^2} + \int {d^4x \over 16\pi^2} \bigg[
        \vsi \si_o \log{\si_o \over e M^2} + {\vsi^2 \over 2} \log{\si_o \over M^2} 
            - \vsi \bigg\{ \Pi({-\grad^2 \over \si_o}) 
            + \Pi({\vsi \over \si_o}) \bigg\} \vsi \bigg].
	\label{e-tr-log-ignoring-deriv-of-cubic-and-higher}
    \eeq
where we ignored cubic and higher powers of $\vsi$ that also carry gradients. We assumed $\vsi \to {\rm const}$ as $|x| \to \infty$ and $\grad \vsi \to 0$ as $|x| \to \infty$. But $\si_o$ need not be the average $\bra \si \ket$. If $\si$ is slowly varying compared to $\si_o$, we ignore terms with $> 2$ derivatives to get
    \beq
    \tr \log{\si -\grad^2 \over M^2} = {\si_o^2 \Om \over 32 \pi^2} 
    	\log{\si_o e^{-{3 \over 2}} \over M^2}
        + \int {d^4x \over 16\pi^2} \bigg[ \vsi \si_o \log{\si_o \over eM^2}
        + {\vsi^2 \over 2} \log{\si_o \over M^2} 
        - {\vsi \grad^2 \vsi \over 6 \si_o} - {\vsi \si_o \over 2} 
        - {3 \vsi^2 \over 4} + {\si^2 \over 2} \log{\si \over \si_o} \bigg].
        \label{e-tr-log-slowly-varying}
    \eeq

\subsection{Scale anomaly $\tr \hat 1 := \zeta(0)$ for general backgrounds}
\label{a-scale-anomaly}

We only got an asymptotic series for $\zeta^{\prime}(0)$ around a constant background, but get a closed-form expression for its scale anomaly. Under a rescaling $\si \mapsto a^{2} \si$, $\zeta(s) \mapsto a^{-2s} \zeta(s)$ so
    \beq
        \zeta^{\prime}(s) \mapsto - 2 \zeta(s) a^{-2s}\log a + a^{-2s} 
        \zeta^{\prime}(s) ~~\implies~~
        \zeta^{\prime}(0) \mapsto \zeta^{\prime}(0) - 2 \zeta(0) \log a
    \eeq
In (\ref{e-expansion-for-zeta-of-s}) most terms are $\propto s$, only $a_{0,1,2}$ contribute to $\zeta(0)$, which is scale-invariant unlike $\zeta^\pr(0)$:
    \beq
        \zeta(0) = {\Om \si_o^2 \over 16 \pi^2} \bigg[
            {\bra a_0 \ket \over 2} - {\bra a_1 \ket \over \si_o} 
            + {\bra a_2 \ket \over \si_o^2} \bigg] 
        = {\Om \si_o^2 \over 16 \pi^2} \bigg[
            \half + {\bra \vsi \ket \over \si_o} 
            + {\bra \vsi^2 \ket \over 2 \si_o^2}
            \bigg].
    \eeq

\section{Original potential $V(\phi^2/N)$ in zeta regularization}
\label{a-back-to-V-legendre-transform}

Subject to the warnings in sec. \ref{s-discussion} about the divergent, scale violating and unphysical nature of $V$ for $\hbar \ne 0$, here we find that for $\zeta$-regularized $W_0(\si_o)$, the {\it finite part} of $V(\eta)$ grows as $\eta^2 / \log \eta$ for large $\eta = \phi^2/N$. We haven't yet determined $V$ for small $\eta$. From (\ref{e-V-to-W-laplace transform}) $e^{-N V(\eta(x))}$ is the inverse Laplace transform of $e^{-N W(\si(x))}$ for each $x$:
	\beq
	(2\pi i)^{-1} \int_{\cal C} d\si~ e^{-(N/2 \hbar) 
		(W(\si) + \si \eta)} &=& e^{-(N/2\hbar)V(\eta)}.
    \eeq
${\cal C}$ is to the right of all singularities of $W(\sigma)$, so it goes from $-i \infty$ to $i \infty$ avoiding ${\bf R}^-$. We must invert the transform for large-$N$, where for constant $\si$ (set $M=1$ in(\ref{e-W_0-for-constant-bkgrnd}))
    \beq
    W(\si) + \eta \si = \eta \si + m^2 \si 
    	- {\hbar \si^2} \log(\tl \la \si)/32 \pi^2  
    {\rm ~~~with~~~~}  \tilde \la 
    	= e^{[-3/2 + 32 \pi^2/\la \hbar ]}.
    \eeq
Splitting into $\Re$ and $\Im$ parts, $\si = u + iv$ and $W(\si) + \eta \si = \vphi + i \psi$. For $\hbar =1$
    \beq
    \vphi &=& (\eta + m^2)u - {(u^2-v^2) 
    	\log{\tl \la \sqrt{u^2 + v^2}} \over 32 \pi^2}
        + {uv \arctan(v/u) \over 16 \pi^2} \cr
    \psi &=& (\eta + m^2)v - {(u^2-v^2) \arctan(v/u) \over 32
    \pi^2} - {uv \log(\tl \la \sqrt{u^2 + v^2}) \over 16
    \pi^2}.
    \eeq
$\vphi \to \infty$ as $v \to \pm \infty$ for all $u \geq 0$. So the integrand $\to 0$ along the lines $u \pm i \infty$ $\forall ~ u$. Thus the end-points of ${\cal C}$ can be moved to $\pm i \infty + u_{\pm}$ for any real $u_{\pm}$ without altering the integral. The strategy for estimating such integrals is as follows \cite{bender-orszag}. $W + \eta \si$ is in general complex on ${\cal C}$. Its $\Im$-part $\psi$ will lead to a highly oscillatory integral as $N \to \infty$ and make it difficult to estimate. The trick is to use analyticity of $W + \eta \si$ to deform $\cal C$ to a (union of) contour(s) on which $\psi$ is constant or where the integrand vanishes. If ${\cal C}$ is a single such contour,
	\beq
    \int_{\cal C} \fr{d\si}{2\pi i} e^{-(N/2 \hbar) (W(\si) + \si
    \eta)} = e^{-{N i \over 2 \hbar} \Im(W(\si) + \si \eta)}
    \int_{\cal C} \fr{d\si}{2\pi i} e^{-(N/2 \hbar) \Re(W(\si) + \si
    \eta)}.
    \eeq
The $N\to \infty$ asymptotics are determined by local minima of $\vphi$ on ${\cal C}$. $\vphi \to \infty$ at end points of ${\cal C}$, so local minima occur at interior points of ${\cal C}$ where directional derivatives of $\vphi,\psi$ vanish along $\cal C$. Since $\vphi + i \psi$ is analytic, local minima of $\vphi$ are saddle points, $\pdr_\si(W + \si \eta)=0$. Not all saddle points may lie on ${\cal C}$. Those that do not, will not contribute to the asymptotics. Saddle points at which $\vphi$ is not a local minimum on $\cal C$ also do not contribute to the asymptotics. Suppose $\si_s(\eta)$ is the only saddle point along ${\cal C}$, then $\vphi$ is a local minimum at $\si_s$. The integrand attains a maximum along ${\cal C}$ at $\si_s$ and decays exponentially in either direction. We approximate $\cal C$ by the tangent at $\si_s$ of length $\eps$ on either side and $\vphi(\si)$ by its quadratic Taylor polynomial. Now let $\eps \to \infty$. $\vphi(\si_s)$ gives the leading contribution while the quadratic term in its Taylor series gives a gaussian integral $ \propto 1/\sqrt{N}$. So
    \beq
    e^{-{N \over 2} V(\eta)} &=& e^{-{iN \over 2} \psi(\si_s)}
        e^{-{N \over 2} \vphi(\si_s)}
        (2\pi i)^{-1} \times {\cal O}(1/\sqrt{N}), \cr
    \implies ~~~ V(\eta) &=& \vphi(\si_s) + i \psi(\si_s) 
    	+ {\cal O} (\log N / N).
    \eeq
If there are several $\si_s$ {\em on} $\cal C$ at which $\vphi$ is a local minimum, we add their contributions. If $\si_s \in {\bf R}$, then $\psi(\si_s) = 0$ doesn't contribute. In practice, we find $\si_s$ and then a suitable constant phase $\cal C$ through it. The saddle point condition for $W(\si) + \si \eta$, given $\eta, m^2$ and $\la$ is
    \beq
    {16 \pi^2} (\eta + m^2) / \hbar = \si \log(\tl \la \si
    \sqrt{e}).
    \label{e-saddle-pt-eqn}
    \eeq
Taking $\Im$ and $\Re$ parts $\implies$ a pair of transcendental equations ($\tl \la =1$, $\hbar =1$, $-\pi < \arctan < \pi$)
    \beq
    {v/2} + (v/2) \log{(u^2 + v^2)} &=&
        - u \arctan(v/u) {\rm ~~~~and} \cr
    16 \pi^2 (\eta + m^2) &=& {u / 2} + (u / 2) \log{(u^2 +
        v^2)} - v \arctan(v/u).
    \eeq
We must solve for $\si_s = u + iv \notin {\bf R}^-$. The $1^{\rm st}$ condition $\implies$ $\si_s$ can lie on ${\bf R}^+$ or on a loop (found numerically) in the $(u,v)$ plane around $(0,0)$ symmetric under reflections about either axis\footnote{This follows from the evenness and oddness of the condition in $u$ and $v$ respectively.} and lying within the rectangle\footnote{The limiting values are obtained by solving the $1^{\rm st}$ saddle point condition for small $v$ and $u$ respectively.} $|u| \leq e^{-3/2}, |v| \leq e^{-1/2}$. However, the $2^{\rm nd}$ condition is satisfied on this loop only for a limited range of values of $\eta + m^2$, namely $m_c^2 \geq \eta + m^2 \geq -({32 \pi \sqrt{e}})^{-1}$ for $-e^{-3/2} \leq u \leq 0$ and $-({32 \pi \sqrt{e}})^{-1} \leq \eta + m^2 \leq -m_c^2$ for $0 \leq u \leq e^{-3/2}$ where $m_c^2 = {\hbar e^{-3/2} \over 16 \pi^2 \tl \la}$. For $\eta + m^2$ in this range, $\si_s$'s could occur on the loop as well as on the positive real $\si$ axis and $\vphi$ may not be a minimum at all of them. For now, we set aside the behavior of $V(\eta)$ for small $\eta$ (in units of $M = 1$), i.e. $\eta + m^2 \leq m_c^2$. For $\eta + m^2 \geq m_c^2$ the only possible $\si_s$ are located on the positive real $\si$ axis. In this case, $\si_s$ are given by solving (\ref{e-saddle-pt-eqn}) with $\si = u \in {\bf R}$. Since we assumed $\eta + m^2 > m_c^2$, the lhs $> 0$, and $\exists !$ solution $\si_s$ found recursively ($\tl \eta = 16 \pi^2 (\eta + m^2)/\hbar$)
    \beq
    \si_s = {\tl \eta \over \log(\tl \la \sqrt{e} \si_s)} 
    	= {\tl \eta \over \log\bigg( {\tl \la \sqrt{e} \tl \eta 
		\over \log (\tl \la \sqrt{e} \si_s)} \bigg)}
     = {\tl \eta \over \log(\tl \la \sqrt{e} \tl \eta) 
     	- \log \log (\tl \la \sqrt{e} \si_s)}
     = \cdots
    \eeq
Thus, for large enough $\eta + m^2$, there is a unique saddle point ($\bar \la = 16 \pi^2 \sqrt{e} \tilde \la / \hbar$)
    \beq
    \si_s \to {16 \pi^2 (\eta + m^2) \over 
    	\hbar \log[\bar \la (\eta + m^2)]}
    {\rm ~~ ~~~as~~~~} \eta + m^2 \to \infty.
    \eeq
We numerically verified the existence of a zero phase contour $\cal C$ from $-i\infty$ to $i\infty$, through $\si_s$ with $\vphi$ necessarily a minimum at $\si_s$. Then using (\ref{e-saddle-pt-eqn}),
    \beq
        V(\eta) = W(\si_s(\eta)) + \eta \si_s(\eta) + {\cal O}(\log N/N) \approx 
        \fr{\hbar \si_s^2}{64 \pi^2} + \half \si_s
        (\eta + m^2).
    \eeq
For large $\eta + m^2$ we get (recall that $\bar \la = 16 \pi^2 \sqrt{e} \tilde \la / \hbar$ and $\tl \la = e^{(-3/2 + 32 \pi^2/\la \hbar)}$)
    \beq
        V(\eta) \to {8 \pi^2 \over \hbar} {(\eta + m^2)^2 
        	\over \log[\bar\la (\eta + m^2)]} \bigg[1
        + \ov{2 \log[\bar\la (\eta + m^2)]} \bigg] {\rm ~~ as~~} 
        	\eta + m^2 \to
        \infty
    \eeq
So for fixed $m$ but large $N$ and $\phi^2/N$, in $\zeta$-regularization, the finite part of $V(\phi^2/N)$ grows as $V(\phi^2/N) \sim {(\phi^4/N^2) \over \log(\bar \la \phi^2/M^2 N)}$. A limiting case is $\hbar =0$, where $W_0 = m^2 \si -\si^2/\la$ is finite and $V(\eta) =  (\la /4)(\eta +m^2)^2$ is the quartic in $\phi$.




\begin{thebibliography}{10}

\footnotesize


\bibitem{tHooft-naturalness}
  G.~'t Hooft 1980 in 
{\em Recent developments in gauge theories}, Ed. by G. 't Hooft et
al., (New York : Plenum).

\bibitem{rajeev-non-triv-fp}
  S.~G.~Rajeev 1996
  e-print arXiv:hep-th/9607100.


\bibitem{luscher-weisz}
  M.~Luscher and P.~Weisz
 1987 {\it Nucl.\ Phys.\ B} {\bf 290}, 25 ; 1988 {\bf 295}, 65; 1989 {\bf 318}, 705.

\bibitem{kuti-lin-shen}
  J.~Kuti, L.~Lin and Y.~Shen 1988
  {\it Phys. Rev. Lett.}  {\bf 61}, 678


\bibitem{bender-triviality}
  C.~M.~Bender and H.~F.~Jones 1988
  {\it Phys. Rev. D} {\bf 38} 2526


\bibitem{rigorous-triviality}
  R.~Fernandez, J.~Frohlich and A.~D.~Sokal 1992
{\em Random walks, critical phenomena, and triviality in quantum field theory,}  Texts and monographs in physics, (Berlin: Springer).

\bibitem{wilson-kogut}
  K.~G.~Wilson and J.~B.~Kogut 1974
  {\it Phys. Rept.}  {\bf 12}, 75


\bibitem{pert-unit-bound}
  B.~W.~Lee, C.~Quigg and H.~B.~Thacker 1977
  {\it Phys. Rev. D} {\bf 16}, 1519

\bibitem{triviality-bound}
  R.~F.~Dashen and H.~Neuberger 1983
  {\it Phys. Rev. Lett.}  {\bf 50}, 1897


\bibitem{SUSY-higgs}
  H.~P.~Nilles  1984
  {\it Phys. Rept.}  {\bf 110}, 1

\bibitem{technicolor}
  E.~Farhi and L.~Susskind 1981
  {\it Phys. Rept.}  {\bf 74}, 277

\bibitem{little-higgs}
  N.~Arkani-Hamed, A.~G.~Cohen and H.~Georgi 2001
  {\it Phys. Lett. B} {\bf 513}, 232
  
\bibitem{coleman-weinberg}
  S.~R.~Coleman and E.~Weinberg 1973
  {\it Phys. Rev. D} {\bf 7}, 1888




\bibitem{meissner-nicolai}
  K.~A.~Meissner and H.~Nicolai 2008
  {\it Phys. Lett. B} {\bf 660} 260




\bibitem{Bardeen:1995kv}
  W.~A.~Bardeen 1995
{\em On naturalness in the standard model,}
FERMILAB-CONF-95-391-T.


\bibitem{slavnov}
  A.~A.~Slavnov 2006
  {\it Theor. Math. Phys.}  {\bf 148}, 1159

\bibitem{halpern-huang}
  K.~Halpern and K.~Huang 1996
  {\it Phys. Rev. D} {\bf 53}, 3252;
  H.~Gies 2001
  {\em Phys. Rev. D} {\bf 63}, 065011


\bibitem{zinn-moshe-large-N-review}
  M.~Moshe and J.~Zinn-Justin 2003
  {\it Phys. Rept.}  {\bf 385} 69


\bibitem{Townsend-phi-sixth}
  P.~K.~Townsend 1977
  {\it Nucl. Phys. B} {\bf 118} 199;
 1976 {\it Phys. Rev. D} {\bf 14} 1715;
 1975 {\it Phys. Rev. D} {\bf 12} 2269.


\bibitem{phi-sixth-3d}
  F.~David, D.~A.~Kessler and H.~Neuberger 1985
  {\it Nucl. Phys. B} {\bf 257} 695.

\bibitem{jackiw-eff-action}
  R.~Jackiw 1974
  {\it Phys. Rev. D} {\bf 9}, 1686.


\bibitem{nair-QFT}
V.~P.~Nair 2005 {\em Quantum field theory: A modern perspective}
(New York: Springer).

\bibitem{abramowitz-stegun}
M. Abramowitz and I. A. Stegun 1964
{\em Handbook of Mathematical Functions} 
(New York: Dover).



\bibitem{witten-dyn-susy-brk}
  E.~Witten 1981
  {\it Nucl. Phys.  B} {\bf 188} 513


\bibitem{dorn-otto-zamolodchikov}
  H.~Dorn and H.~J.~Otto 1992
  {\it Phys. Lett.  B} {\bf 291}, 39;
  A.~B.~Zamolodchikov and A.~B.~Zamolodchikov 1996
  {\it Nucl. Phys.  B} {\bf 477}, 577.

\bibitem{liam-sreedhar}
  L.~O'Raifeartaigh, J.~M.~Pawlowski and V.~V.~Sreedhar 1999
  {\it Annals Phys.} {\bf 277}, 117


\bibitem{BMB}
  W.~A.~Bardeen, M.~Moshe and M.~Bander 1984
  {\it Phys. Rev. Lett.}  {\bf 52} 1188


\bibitem{string-dual}
  I.~R.~Klebanov and A.~M.~Polyakov 2002
  {\it Phys. Lett. B} {\bf 550}, 213;
  H.~J.~Schnitzer 2004
  {\it Nucl. Phys.  B} {\bf 695}, 283


\bibitem{richter} B. Richter 2006 (October)
Reference Frame, {\it Physics Today}, 8

\bibitem{atomic-parity-viol} I. B. Khriplovich 1991 {\em Parity Nonconservation in Atomic Phenomena}, (Amsterdam: Gordon and Breach)



\bibitem{bender-PT}
  C.~M.~Bender and S.~Boettcher 1998
  {\it Phys. Rev. Lett.} {\bf 80} 5243



\bibitem{pdg}
  C.~Amsler {\it et al.}  [Particle Data Group] 2008
  Phys.\ Lett.\  B {\bf 667} 1

  
\bibitem{ward-id-sde}
  L.~Akant and G.~S.~Krishnaswami 2007
  {\it JHEP} {\bf 0702}, 073

\bibitem{tHooft-nobbenhuis}
  G.~'t Hooft and S.~Nobbenhuis 2006
  {\it Class. Quant. Grav.} {\bf 23}, 3819


\bibitem{heat-kernel-expansion}
  D.~V.~Vassilevich 2003,
  {\it Phys. Rept.} {\bf 388} 279



\bibitem{bender-orszag}
C. M. Bender and S. A. Orszag 2005 {\em Advanced Mathematical Methods
for Scientists and Engineers: Asymptotic Methods and
Perturbation Theory} (New York: Springer).






\end{thebibliography}
\end{document}